\documentclass{article}

\usepackage[nonatbib, final]{neurips_2019}

\usepackage[utf8]{inputenc} 
\usepackage[T1]{fontenc}    
\usepackage{hyperref}       
\usepackage{url}            
\usepackage{booktabs}       
\usepackage{amsfonts}       
\usepackage{nicefrac}       
\usepackage{microtype}      
\usepackage{amsfonts}
\usepackage{amsmath}
\usepackage{amssymb}
\usepackage{graphicx}
\usepackage{wrapfig}
\usepackage{mathptmx}

\providecommand{\mbf}[1]{\mathbf{#1}}

\providecommand{\der}[2]{\frac{\partial #1}{\partial #2}}

\title{Intrusive deconvolutional neural networks\\ for enhancing PIC/FLIP solutions}

\author{
  Y. van Halder, \ B.Sanderse\\
  Scientific Computing Group\\
  Centrum Wiskunde \& Informatica (CWI) \\
  P.O. Box 94079, 1090 GB, Amsterdam, the Netherlands \\
  \texttt{yvanhalder@gmail.com} \\ 
  \And
  B. Koren \\
  Center for Analysis, Scientific Computing and Applications \\
  Eindhoven University of Technology \\
  P.O. Box 513, 5600 MB, Eindhoven, the Netherlands
}

\begin{document}

\maketitle

\begin{abstract}
Traditional fluid flow predictions require large computational resources. Despite recent progress in parallel and GPU computing, the ability to run fluid flow predictions in real-time is often infeasible. Recently developed machine learning approaches, which are trained on high-fidelity data, perform unsatisfactorily outside the training set and remove the ability of utilising legacy codes after training. We propose a novel methodology that uses a deep learning approach that can be used within a low-fidelity fluid flow solver to significantly increase the accuracy of the low-fidelity simulations. The resulting solver enables accurate while reducing computational times up to 100 times. The deep neural network is trained on a combination of low- and high-fidelity data, and the resulting solver is referred to as a multi-fidelity solver. The proposed methodology is demonstrated by means of enhancing a fluid flow simulator, known as PIC/FLIP, which is a popular fluid flow simulator in the field of computer generated imagery.
\end{abstract}

\section{Introduction}
The huge increase in computational power over the last decades has resulted in an increasing interest in real-time computations for physics problems. These real-time computations may be used for monitoring a physical asset or real-time optimisation of a dynamic process. In many industrial processes, such as coastal engineering \cite{barreiro_smoothed_2013}, vehicle design \cite{guo_convolutional_2016}, computer generated imagery \cite{mcadams_detail_2009,frost_moana:_2017, stomakhin_material_2013, angelidis_controllable_2006}, fluid flow predictions are essential. In fluid dynamics, a distinction can be made between low-fidelity and high-fidelity simulations. Low-fidelity simulations are computationally cheap to perform, but have limited accuracy, as they do not capture all the underlying physics or do not resolve all scales. On the other hand, high-fidelity simulations incorporate all the relevant physical phenomena and corresponding scales, but may require a tremendous amount of computational resources. This makes it difficult to this date to perform accurate fluid simulations in real-time.

In order to address the problem of computational expense, various types of methods have been introduced. A commonly used method is Principal Component Analysis \cite{ravindran_reduced-order_2000, audouze_reduced-order_2009}, which provides the desired speed-up by transforming the dynamics of the fluid flow simulation to operations on linear combinations of snapshots, therefore restricting the richness of the dynamics. Recently developed data-driven methods use machine learning algorithms \cite{ladicky_data-driven_2015,ladicky_physicsforests:_2017, guo_convolutional_2016}, such as regression forests and neural networks, that are trained on a large set of high-fidelity fluid flow simulation data. After training, these approaches are able to simulate fluid flows in real-time. However, they often perform unsatisfactorily outside the training set resulting in unrealistic fluid flow simulations. This extrapolation problem is a weakness of almost all machine learning approaches but can be partially overcome by supplying the machine learning algorithm with more information about the underlying physics problem during training \cite{ladicky_data-driven_2015}. Furthermore, these recent deep learning methods have not yet been combined with existing fluid flow solvers, which have been developed and validated over many years.

We propose a novel method that \textit{enhances a low-fidelity fluid flow solver by training a deep neural network that maps values from a low-resolution computational grid to a high-resolution computational grid. This neural network is then used intrusively to enhance low-fidelity simulations at each time step}. The input and output of the neural network are the current state of the low-fidelity simulation and the approximate high-fidelity current state, respectively. Compared to data-driven approaches that are purely trained on high-fidelity data, the crucial advantage of our model+data-driven approach is that the low-fidelity simulation is acting as a preconditioner, as it already comprises parts of the physics involved in the problem. Our proposed multi-fidelity approach is able to significantly enhance the accuracy of low-fidelity fluid flow predictions, even outside the training set.

In this work, we focus on enhancing a so-called Particle In Cell/Fluid Implicit Particle (PIC/FLIP) solver \cite{zhu_animating_2005}. The PIC/FLIP solver is a popular fluid flow solver in the field of computer generated imagery (CGI) \cite{bridson_fluid_2015, jiang_affine_2015}, because of its straightforward parallel implementation. The solver evolves a set of particles (representing the fluid) over time by computing fluid flow velocities on a staggered Cartesian grid where the fidelity of the simulation is determined by the resolution of the grid and the total number of particles. A parallel implementation of the PIC/FLIP solver is able to simulate millions of particles on a coarse underlying grid in real-time. The capability to simulate a fluid flow in real-time is promising, but as these solutions are in general computed using a coarse grid, it often results in inaccurate fluid flow simulations. Our goal is to \textit{increase the accuracy of 3D low-fidelity PIC/FLIP simulations at each time step, by incorporating into the solver: a deep neural network, which is trained on high-fidelity data.} This allows for accurate simulations in real-time. We focus on enhancing PIC/FLIP sloshing simulations in a rectangular tank, as it is commonly encountered in CGI applications. This test case comprises a wide range of fluid mechanics phenomena.

To see where the neural networks fit into the PIC/FLIP framework, we first discuss the PIC/FLIP solver in detail in section 2. Section 3 discusses our multi-fidelity approach in detail. Lastly, section 4 studies the trained neural network in more detail, shows numerical results of our approach in enhancing a 3D sloshing simulation, and shows how the trained neural network generalises outside the training set.

\section{PIC/FLIP for Simulating Fluid Flows in 3D}
\label{sec:fluidsolver}
\label{sec:PICFLIP}
In this section we discuss the fluid flow solver in more detail. The governing equations for incompressible fluid flow are the Navier-Stokes equations:
\begin{subequations}
\begin{align}
\nabla\cdot \mbf{u} &= 0 \label{eqNS:incompressible}\ ,\\
\der{\mbf{u}}{t} + (\mbf{u}\cdot\nabla)\mbf{u} &= -\frac{\nabla p}{\rho} + \mbf{F} + \nu \nabla^2\mbf{u}\label{eqNS:momentum}\ ,
\end{align}
\end{subequations}
where $\mbf{u}$ is the velocity field, $p$ the pressure, $\rho$ the density, $\mbf{F}$ the body-force vector, and $\nu$ the kinematic viscosity. The first equation \eqref{eqNS:incompressible} is known as the incompressibility condition, while the second set of equations \eqref{eqNS:momentum} is known as the momentum equation. When solving these equations, we can resort to grid-based methods or particle-based methods. In this work we use a PIC/FLIP solver for predicting single-phase free-surface flows, as it allows for real-time computation when implemented efficiently on a GPU. This numerical method is a combination of a grid-based method and a particle-based method, and is often used to simulate fluid motions in movies or computer games due to its easy parallel implementation. The particles represent the fluid, and the positions of these particles are evolved over time by using velocity values that are computed on a staggered grid. In this section we discuss the individual steps in the solver in more detail, as they determine the architecture of our neural network, which is discussed in section \ref{sec:method}.

The PIC/FLIP framework comprises two stages; 1) initialisation of grid and particles, 2) time-stepping loop.
\subsection{Initialisation}
\label{sec:init}
\begin{figure}[!h]
\centering
\includegraphics[width = \textwidth]{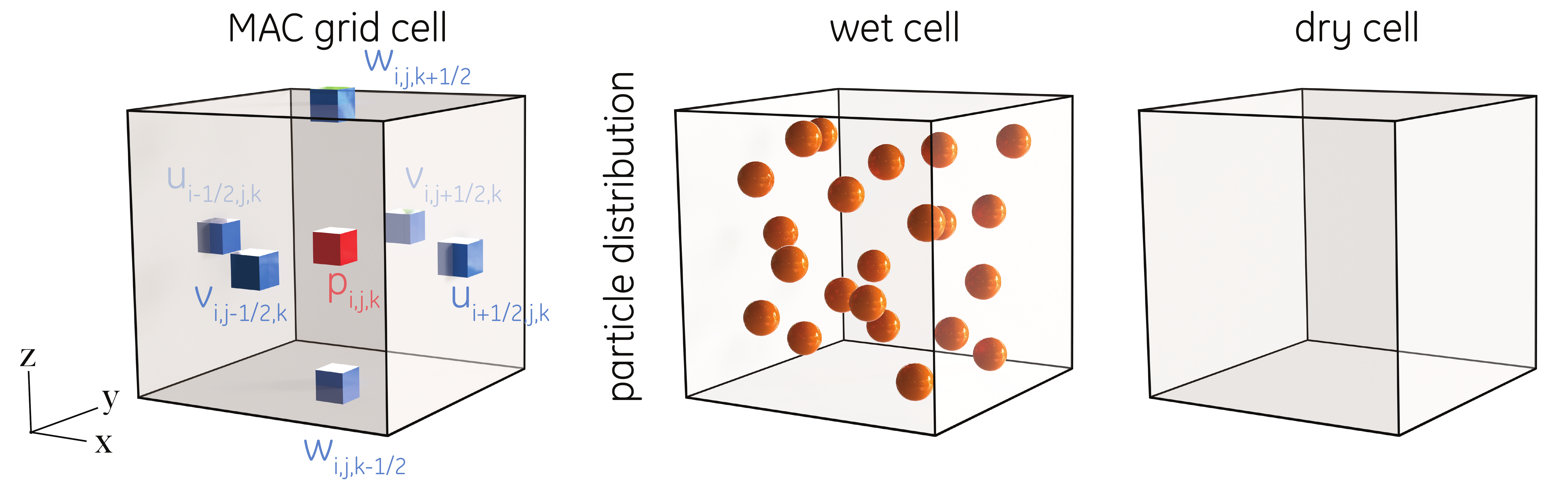}
\caption{\label{fig:PICFLIP_initialisation} The staggered grid quantities for the PIC/FLIP solver.}
\end{figure}

The initialisation stage is shown schematically in figure \ref{fig:PICFLIP_initialisation}.
\\ 

\textit{MAC staggered grid}\\
A Marker And Cell (MAC) staggered grid is used, which defines the velocity vector $\mbf{u}=(u, v, w)$ at the corresponding cell faces (blue cubes), while the pressure is defined at the centre of the cell (red cube). The main reason for using a staggered grid is to remove the odd-even coupling of velocity and pressure, which may occur when using a collocated grid where velocity and pressure are defined at the same locations. The staggered grid is characterised by the domain dimensions $(L_x, L_y, L_z)$ and the number of cells in each of the three coordinate directions $(N_x, N_y, N_z)$, which results in cells with dimensions $(\Delta x, \Delta y, \Delta z) = (\frac{L_x}{N_x}, \frac{L_y}{N_y}, \frac{L_z}{N_z})$. In the remainder of this paper, we assume that the number of grid cells is chosen such that $\Delta x = \Delta y = \Delta z$, and this uniform cell width is denoted as $\Delta s$. The indexing of the pressure and face velocities is shown in figure \ref{fig:PICFLIP_initialisation}. The indices $i, j, k$ correspond to the cell number in $x, y, z$-direction, respectively, where the first cell in each direction has an index 0. To clarify, $u_{-\frac{1}{2}, j, k}$ and $u_{N_x-\frac{1}{2}, j, k}$ denote the $x$-component of the velocity at the left and right boundaries of the domain, respectively. \\

\begin{figure}[!h]
\centering
\includegraphics[width = \textwidth]{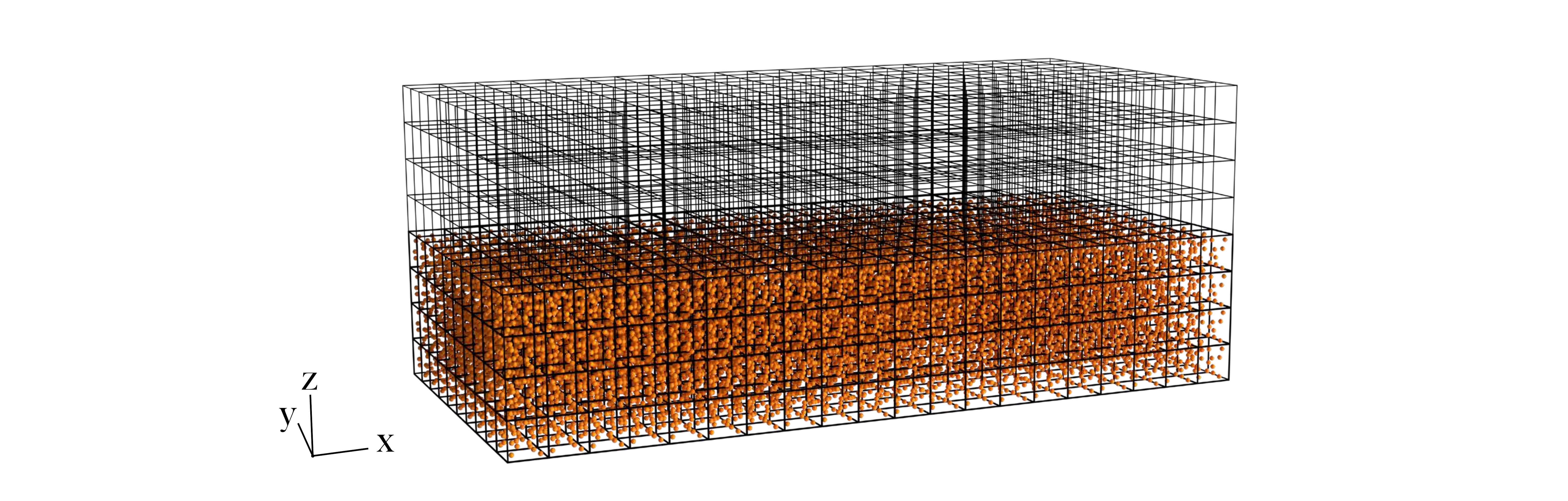}
\caption{\label{fig:PICFLIP_initialisation2} Grid and particle initialisation with $(N_x, N_y, N_z)=(20, 8, 8)$. The top half of the grid comprises dry cells while the cells in the bottom half are filled with particles.}
\end{figure}

\textit{Particle initialisation}\\
After the grid has been initialised, we specify which of the cells need to be filled with particles, i.e., which cells are wet, corresponding to the initial state of the fluid. Following the implementation in \cite{bridson_fluid_2015} we place 8 randomly placed particles within each wet cell and set the initial velocity of these particles to zero. We denote the total number of particles by $N_p$, and by $\mbf{p}^{(i)}=(p^{(i)}_{x}, p^{(i)}_{y}, p^{(i)}_{z})$ and $\mbf{p_{\mbf{u}}}^{(i)}=(p^{(i)}_{u}, p^{(i)}_{v}, p^{(i)}_{w})$ we denote the position and velocity of the $i$-th particle, respectively. An example of an initialisation of grid cells and particles is shown in figure \ref{fig:PICFLIP_initialisation2}.

\subsection{Time-stepping loop}
The time-stepping loop comprises 8 steps, which are performed in a sequential loop to advance the current solution at time $t$ to the new time-level $t+\Delta t$, where $\Delta t$ denotes the time step. These 8 steps are detailed below.

\subsubsection{Particle to grid transfer}
\begin{figure}[!h]
\centering
\includegraphics[width = \textwidth]{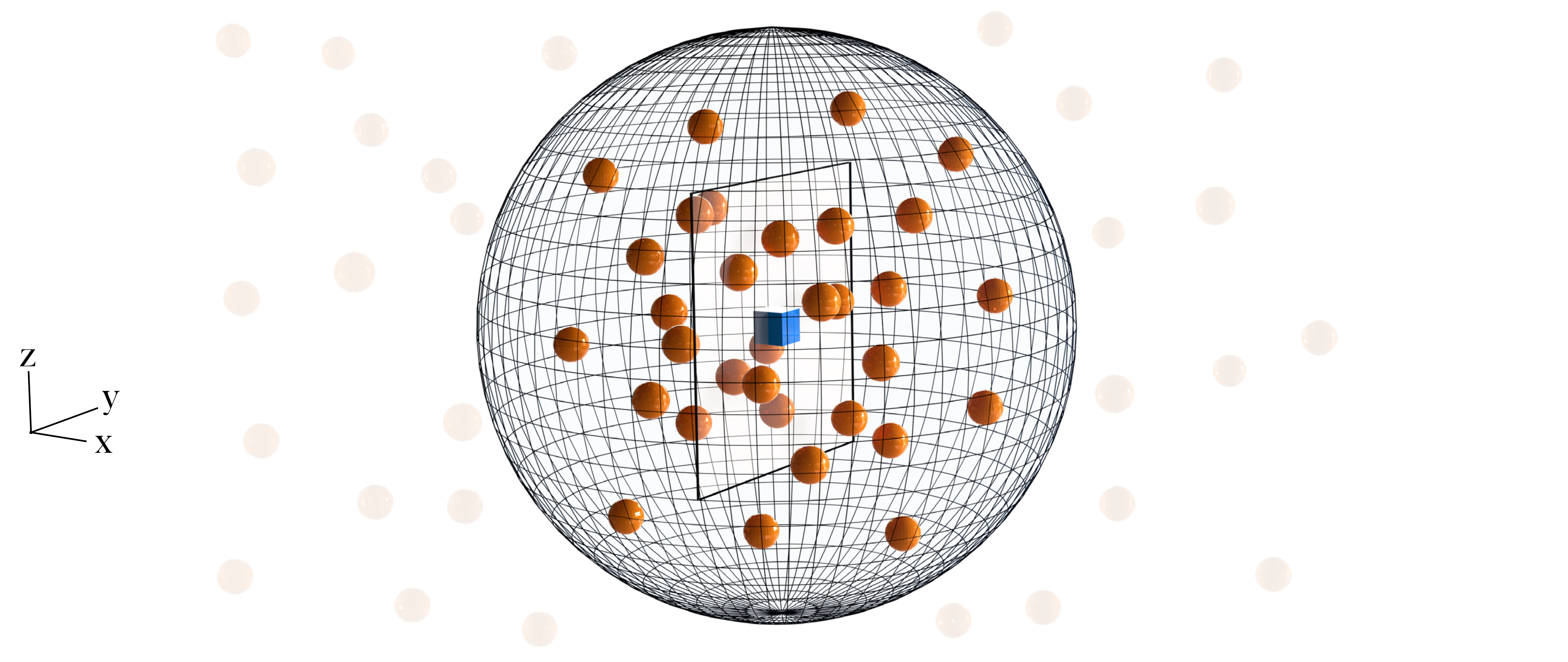}
\caption{\label{fig:PICFLIP_P2G} Particles that are used for computing one face velocity.}
\end{figure}

A weighted average of particle velocities is used to compute the velocities at the grid-cell faces. In order to compute the velocity at a given face centre, particles that lie within a sphere with a $2\Delta s$ radius, centred around this face centre, are used for computing the new velocity value. Consequently, the new face centre velocities are given by:
\begin{subequations}
\begin{align}
u_{i+\frac{1}{2}, j, k} &= \frac{\sum_{i=1}^{N_p} p^{(i)}_u h(\mbf{p}^{(i)} - \mbf{x}_{i+\frac{1}{2}, j, k})}{\sum_{i=1}^{N_p}h(\mbf{p}^{(i)} - \mbf{x}_{i+\frac{1}{2}, j, k})}\ ,\\
v_{i, j+\frac{1}{2}, k} &= \frac{\sum_{i=1}^{N_p} p^{(i)}_v h(\mbf{p}^{(i)} - \mbf{x}_{i, j+\frac{1}{2}, k})}{\sum_{i=1}^{N_p}h(\mbf{p}^{(i)} - \mbf{x}_{i, j+\frac{1}{2}, k})}\ ,\label{eqPICFLIP:kernel}\\
w_{i, j, k+\frac{1}{2}} &= \frac{\sum_{i=1}^{N_p} p^{(i)}_w h(\mbf{p}^{(i)} - \mbf{x}_{i, j, k+\frac{1}{2}})}{\sum_{i=1}^{N_p}h(\mbf{p}^{(i)} - \mbf{x}_{i, j, k+\frac{1}{2}})}\ ,
\end{align}
\end{subequations}
where the vectors $\mbf{x}$ correspond to the locations of the face centres, and where $h$ is the so-called kernel. The choice of kernel $h$ depends on the application. In general the commonly used cubic kernel is considered to be robust and is therefore employed in the remainder of this paper\cite{bridson_fluid_2015}:
\begin{align}
k(\mbf{r}) \left\lbrace \begin{array}{l l}
(4\Delta s^2 - \|\mbf{r}\|_2^2)^3 &,\ \|\mbf{r}\|_2\leq 2\Delta s\ ,\\
0&,\ \text{otherwise}\ .
\end{array}\right.
\end{align}
The particles that are used for the grid transfer for a single face velocity are schematically shown in figure \ref{fig:PICFLIP_P2G}.
\subsubsection{Store velocities}
After the particle velocities have been transferred to the grid, the resulting face velocities are copied for later use, i.e., we define:
\begin{equation}
u^*_{i+\frac{1}{2}, j, k} = u_{i+\frac{1}{2}, j, k},\quad v^*_{i, j+\frac{1}{2}, k} = v_{i, j+\frac{1}{2}, k},\quad w^*_{i, j, k+\frac{1}{2}} = w_{i, j, k+\frac{1}{2}}\ .
\label{eqPICFLIP:originalvelocities} 
\end{equation}
The copied velocities remain unaltered for the rest of the time step.

\subsubsection{Add body forces}
The external forces, such as gravity, are added to the velocity field by a simple forward Euler time-integration:
\begin{subequations}
\begin{align}
u_{i+\frac{1}{2}, j, k} &= u_{i+\frac{1}{2}, j, k} + \Delta t F_x(\mbf{x}_{i+\frac{1}{2}, j, k})\ ,\label{eq:face_intermediate}\\
v_{i, j+\frac{1}{2}, k} &= v_{i, j+\frac{1}{2}, k} + \Delta t F_y(\mbf{x}_{i, j+\frac{1}{2}, k})\ ,\\ 
w_{i, j, k+\frac{1}{2}} &= w_{i, j, k+\frac{1}{2}} + \Delta t F_z(\mbf{x}_{i, j, k+\frac{1}{2}})\ ,\label{eq:face_intermediate2}
\end{align}
\end{subequations}
where $F_x,\ F_y$ and $F_z$ are the components of the body force acting on the fluid in each coordinate direction, calculated at the corresponding face centres. A more accurate time-integration scheme may be used, but forward Euler is very easy to implement, and is therefore considered as the standard in PIC/FLIP \cite{zhu_animating_2005, bridson_fluid_2015}. When $\Delta s$ is set, we need to choose $\Delta t$ accordingly such that the forward Euler time integration remains stable, which is often not an issue when using PIC/FLIP due to the often coarse computational grid.

\subsubsection{Enforce boundary conditions}
We assume that the domain boundaries and obstacles are solid, and therefore solid-wall boundary conditions are imposed at these locations, which state that fluid is not allowed to flow out of the domain. As a result, the boundary condition is given by:
\begin{equation}
\mbf{u}\cdot \mbf{n} = \mbf{u}_{\text{boundary}}\cdot \mbf{n}\ ,
\end{equation}
where $\mbf{n}$ is the normal vector pointing outside the domain. Furthermore, this boundary condition can be simplified when using a staggered grid, where the faces intersect with the domain boundaries:
\begin{subequations}
\begin{align}
u_{-\frac{1}{2}, j, k} &= u_{\text{boundary},0},\ \ u_{N_x-\frac{1}{2}, j, k} = u_{\text{boundary},L_x}\ ,\\
v_{i, -\frac{1}{2}, k} &= v_{\text{boundary},0},\ \ v_{i, N_y-\frac{1}{2}, k} = v_{\text{boundary},L_y}\ ,\\
w_{i, j, -\frac{1}{2}} &= w_{\text{boundary},0},\ \ w_{i, j, N_z-\frac{1}{2}} = w_{\text{boundary},L_z}\ ,
\end{align}
\end{subequations}
where we assumed that the domain does not deform and the boundaries have a uniform velocity, i.e., $u_{\text{boundary},0} = u_{\text{boundary},L_x},\ v_{\text{boundary},0} = v_{\text{boundary},L_y},\ w_{\text{boundary},0} = w_{\text{boundary},L_z}$.

\subsubsection{Determine which cells contain particles}
The cells that contain at least one particle are referred to as wet cells. We define a new quantity $m_{i, j, k}$:
\begin{equation}
m_{i, j, k} \left\lbrace \begin{array}{l l}
1 &,\ \text{if cell } (i, j, k) \text{ contains at least one particle}\ ,\\
0&,\ \text{otherwise}\ .
\end{array}\right.
\end{equation}

\subsubsection{Pressure correction}
The PIC/FLIP method simulates incompressible fluid flow, which means that the velocities at the face centres of all $i,j,k$ have to satisfy the following discrete version of incompressibility constraint \eqref{eqNS:incompressible}:
\begin{equation}
(\nabla\cdot\mbf{u})_{i, j, k} \rightarrow \frac{u_{i+\frac{1}{2}, j, k} - u_{i-\frac{1}{2}, j, k}}{\Delta s} + \frac{v_{i, j+\frac{1}{2}, k} - v_{i, j-\frac{1}{2}, k}}{\Delta s} + \frac{w_{i, j, k+\frac{1}{2}} - w_{i, j, k-\frac{1}{2}}}{\Delta s} = 0\ ,
\label{eqPICFLIP:discreteincompressible}
\end{equation}
which states that the central second-order accurate divergence, computed at each cell centre, should be zero. In general, the face velocities from \eqref{eq:face_intermediate}-\eqref{eq:face_intermediate2} do not satisfy this constraint. Therefore, the velocities are corrected by subtracting a pressure gradient. In order to calculate this pressure gradient, we solve the pressure equation in the cells that are wet ($m_{i, j, k}=1$):
\begin{equation}
(\nabla^2 p)_{i, j, k} = \frac{1}{\Delta t}(\nabla\cdot\mbf{u})_{i, j, k}\ ,
\end{equation}
where
\begin{equation}
(\nabla^2 p)_{i, j, k} \rightarrow \frac{p_{i+1, j, k} + p_{i-1, j, k}}{\Delta s^2} + \frac{p_{i, j+1, k} + p_{i, j-1, k}}{\Delta s^2} + \frac{p_{i, j, k+1} + p_{i, j, k-1}}{\Delta s^2} - \frac{6p_{i, j, k}}{\Delta s^2}\ ,
\label{eqPICFLIP:pressureequation}
\end{equation}
and set $p_{i, j, k} = 0$ in dry cells. Notice that near the domain boundaries or obstacles we require the pressures located outside the domain, which are not defined, and a different formulation is necessary. To enforce that there is no flow through solid boundaries, we employ the commonly used homogeneous Neumann boundary condition for the pressure at these boundaries. Even though the homogeneous Neumann boundary condition may translate to a non-zero tangential velocity component along the boundaries, it is the easiest to implement while still enforcing that no fluid flows outside the domain. Other boundary conditions, e.g., strict no-slip boundary conditions, are not straightforward to implement on a staggered grid and are not considered in this paper. The homogeneous Neumann boundary conditions set the pressure coefficient of the cells outside the domain or inside an obstacle to zero and the central pressure coefficient 6 in \eqref{eqPICFLIP:pressureequation} is decreased by the number of cells outside the domain or inside an obstacle, which translated to a pressure gradient that is zero on the boundary \cite{bridson_fluid_2015}. All this results in a sparse linear system of the form:
\begin{equation}
A\left(\begin{array}{c}
p_{0, 0, 0}\\
p_{1, 0, 0}\\
\vdots\\
p_{N_x-1, N_y-1, N_z-1}
\end{array}\right) = \frac{1}{\Delta t}\left(\begin{array}{c}(\nabla\cdot\mbf{u})_{0, 0, 0}\\
(\nabla\cdot\mbf{u})_{1, 0, 0}\\
\vdots\\
(\nabla\cdot\mbf{u})_{N_x-1, N_y-1, N_z-1}\end{array}\right)\ ,
\label{eq:pressureequation}
\end{equation}
where $A$ is the pressure Poisson matrix. This system is solved using the preconditioned conjugate gradient method, as it does not require the explicit storage of the coefficient matrix $A$, and still exhibits relatively fast convergence for symmetric matrices $A$ when compared to Jacobi iteration.

After the pressures have been computed, we correct the face velocities, that do not coincide with a domain boundary, as follows:
\begin{subequations}
\begin{align}
u_{i+\frac{1}{2}, j, k} &= u_{i+\frac{1}{2}, j, k} - \frac{\Delta t}{\Delta s}(p_{i+1, j, k} - p_{i, j, k})\ ,\\
v_{i, j+\frac{1}{2}, k} &= v_{i, j+\frac{1}{2}, k} - \frac{\Delta t}{\Delta s}(p_{i, j+1, k} - p_{i, j, k})\ ,\label{eq:incompressible_velocities}\\ 
w_{i, j, k+\frac{1}{2}} &= w_{i, j, k+\frac{1}{2}} - \frac{\Delta t}{\Delta s}(p_{i, j, k+1} - p_{i, j, k})\ ,
\end{align}
\end{subequations}
which now satisfy the constraint \eqref{eqPICFLIP:discreteincompressible}.

\subsubsection{Grid to particle transfer}
Once the grid velocities are known, they have to be transferred back to the particles in order to advect the particles. To calculate the new particle velocities, we use \textit{trilinear interpolation}. We will discuss how to compute the $u-$component of the particle velocity $p_u^{(i)}$, and the remaining two components can be computed in a similar fashion.

\begin{figure}[!h]
\centering
\includegraphics[width = \textwidth]{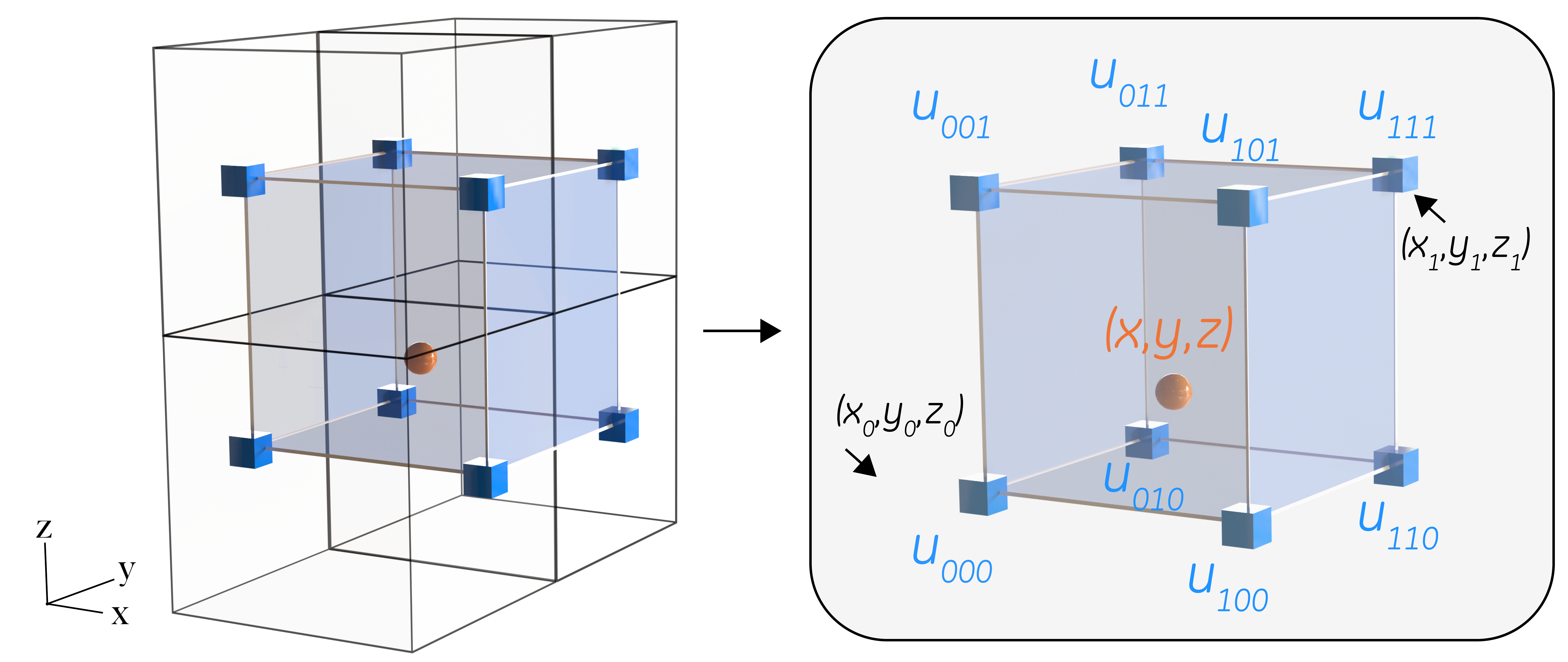}
\caption{\label{fig:PICFLIP_G2P} The face velocities that are used for the trilinear interpolation procedure to calculate the $u-$component of the particle velocity.}
\end{figure}

The face centre velocities that are used for the trilinear interpolation procedure for computing the $u-$component of the velocity $p_u^{(i)}$ for a single particle are shown in figure \ref{fig:PICFLIP_G2P}.
\\
The $u-$component of the velocity is now calculated as:
\begin{subequations}
\begin{gather}
x^* = \frac{x-x_0}{x_1-x_0},\quad y^* = \frac{y-y_0}{y_1-y_0},\quad z^* = \frac{z-z_0}{z_1-z_0}\ ,\\
\nonumber\\
u_{00} = u_{000}(1-x^*) + u_{100} x^*,\quad u_{10} = u_{001}(1-x^*) + u_{101} x^*\nonumber\ ,\\
u_{01} = u_{010}(1-x^*) + u_{110} x^*,\quad u_{11} = u_{011}(1-x^*) + u_{111} x^*\ ,\\
\nonumber\\
u_{0} = u_{00}(1-y^*) + u_{01} y^*\nonumber\ ,\\
u_{1} = u_{10}(1-y^*) + u_{11} y^*\ ,\\
\nonumber\\
p_u^{(i)} = u_{0}(1-z^*) + u_{1} z^*\ .
\end{gather}
\end{subequations}
This trilinear interpolation is performed two times to perform the PIC/FLIP velocity update; with the new grid velocities $u_{i+\frac{1}{2}, j, k}$ from \eqref{eq:incompressible_velocities}, and the old grid velocities $u^*_{i+\frac{1}{2}, j, k}$ from \eqref{eqPICFLIP:originalvelocities}. As a result, two particle velocities are obtained $p_{\text{new}, u}^{(i)}$ and $p_{\text{old}, u}^{(i)}$. The new particle velocity is calculated as follows:
\begin{equation}
p_u^{(i)} \leftarrow (1-f) (p_u^{(i)} - p_{\text{old}, u}^{(i)}) + p_{\text{new}, u}^{(i)} \ ,
\end{equation}
where $p_u^{(i)}$ is the particle velocity from the previous time step, and $f\in[0, 1]$ the so-called PICness parameter. If $f=1$, then the velocity update is known as the PIC update, while setting $f=0$ corresponds to the FLIP update. PIC is known for its good stability properties, but suffers from severe numerical diffusion. On the other hand, FLIP does not suffer from numerical diffusion, but can become unstable. Therefore, a weighted average between PIC and FLIP is taken, $f\in(0, 1)$, to obtain a stable, but less diffusive solution. Notice that we did not include physical diffusion induced by the viscosity term in the method so far. In PIC/FLIP methods, the physical diffusion term in the Navier-Stokes equations is approximated by tweaking the numerical viscosity by choosing a specific value for the PICness parameter $f$, namely, $f$ can be related to the kinematic viscosity as:
\begin{equation}
f = \max\left\lbrace 1, \frac{6\Delta t \nu}{\Delta s ^2}\right\rbrace\ ,
\end{equation}
and the value for $\nu$ should be set to the desired kinematic viscosity accordingly. However, it is common practice to use $f=0.99$, which yields good results for water flow \cite{bridson_fluid_2015} and is used in the remainder of this paper. Physical diffusion by viscosity can be implemented in a PIC/FLIP solver but also requires no-slip boundary conditions and is not considered in this paper for ease of implementation.

\subsubsection{Advect particles}
A suitable time-integrator is used to advect the particles. A good trade-off between computational expense and accuracy is the second-order accurate Runge-Kutta method (RK2) which mimics the midpoint rule. The procedure for advecting the particles is given by:
\begin{enumerate}
\item Advect the particles for the duration of a half time step: $\mbf{p}^{(i)}_{\text{halfway}}=\mbf{p}^{(i)} + \frac{\Delta t}{2}\mbf{p}_{\mbf{u}}^{(i)}$,
\item Compute new particle velocities using trilinear interpolation, using the grid velocities $u_{i+\frac{1}{2}, j, k}$ to obtain $\mbf{p}_{\text{halfway},\mbf{u}}^{(i)}$,
\item Compute new particle positions as $\mbf{p}^{(i)} = \mbf{p}^{(i)} + \Delta t \mbf{p}_{\text{halfway},\mbf{u}}^{(i)}$.
\end{enumerate}
After computing new particle positions $\mbf{p}^{(i)}_{\text{halfway}}$ and $\mbf{p}^{(i)}$, it may appear that some particles have crossed the boundary of the domain. Particles that crossed the domain boundary are reflected back into the domain to prevent particles leaving the domain \cite{bridson_fluid_2015}.

\subsection{PIC/FLIP for fluid sloshing}
The main focus of this paper is to enhance low-fidelity PIC/FLIP sloshing simulations. These sloshing simulations can be performed in a number of ways and we choose the one that is easiest to implement. In order to induce fluid sloshing, the gravity vector is rotated along the $x$-axis and $y$-axis independently. To clarify, the rotation of the gravity vector enters the fluid solver in the body force as $\mbf{F}=\mbf{g}=9.81(\sin(\psi)\cos(\phi),\sin(\phi), -\cos(\psi)\cos(\phi))$, where $\psi$ and $\phi$ are the rotation angles along the $x$ and $y$ axis, respectively. Rotating the gravity vector mimics rotation of the tank without the need to complicate the boundary conditions of the tank boundary to incorporate rotational accelerations.

\subsection{Accuracy of PIC/FLIP}
The accuracy of a PIC/FLIP simulation is determined by the number of particles $N_p$ and the grid resolution $\Delta s$. As the particles do not react with each other, the steps that involve particles are easily implemented on a GPU. The number of particles can be chosen to be large (millions) while still being able to run in real-time. Most computational work occurs in the computation on the grid, especially in the solution of equation \eqref{eq:pressureequation} to compute the pressures at the cell centres. It is not straightforward to efficiently implement these computations on a GPU. Solving the pressure equation is done in $O(N_c \log(N_c))$ time, where $N_c$ is the total number of grid cells. Decreasing the grid resolution by a factor 2 in each coordinate direction, i.e., $\Delta s \rightarrow 2\Delta s$, results in a pressure solve that is approximately 10 times faster. As a result, when choosing a coarse grid, we can still easily simulate up to millions of particles in real-time, but computing solutions on a coarse grid results in a significant drop in accuracy of the particle positions when compared to simulating flows on a fine grid with the same number of particles. This drop in accuracy is mainly caused by the inaccurate incompressible velocity field that is computed on a the coarse grid, and the trilinear interpolation that is used to transfer these grid velocities to the particles. For instance, the error in the trilinear interpolation is composed of three linear interpolation errors $R_{\text{int}}$, which are bounded by \cite{phillips_interpolation_2003}
\begin{equation}
R_{\text{int}} \leq C\Delta s^2\ ,
\label{eq:interpolationerror}
\end{equation}
where $C$ is a constant depending on the smoothness of the velocity field that is transferred to the particles. Equation \eqref{eq:interpolationerror} expresses that the error increases quadratically when coarsening the grid. Insufficiently accurate solutions may quickly arise when coarsening the computational grid. This quadratic scaling of errors holds for other grid computations as well (divergence computation \eqref{eqPICFLIP:discreteincompressible}, pressure computation \eqref{eq:pressureequation}). 

\section{Deep learning for enhancing low-fidelity fluid flow predictions}
\label{sec:method}
\begin{figure}[hbt!]
\includegraphics[width = \textwidth]{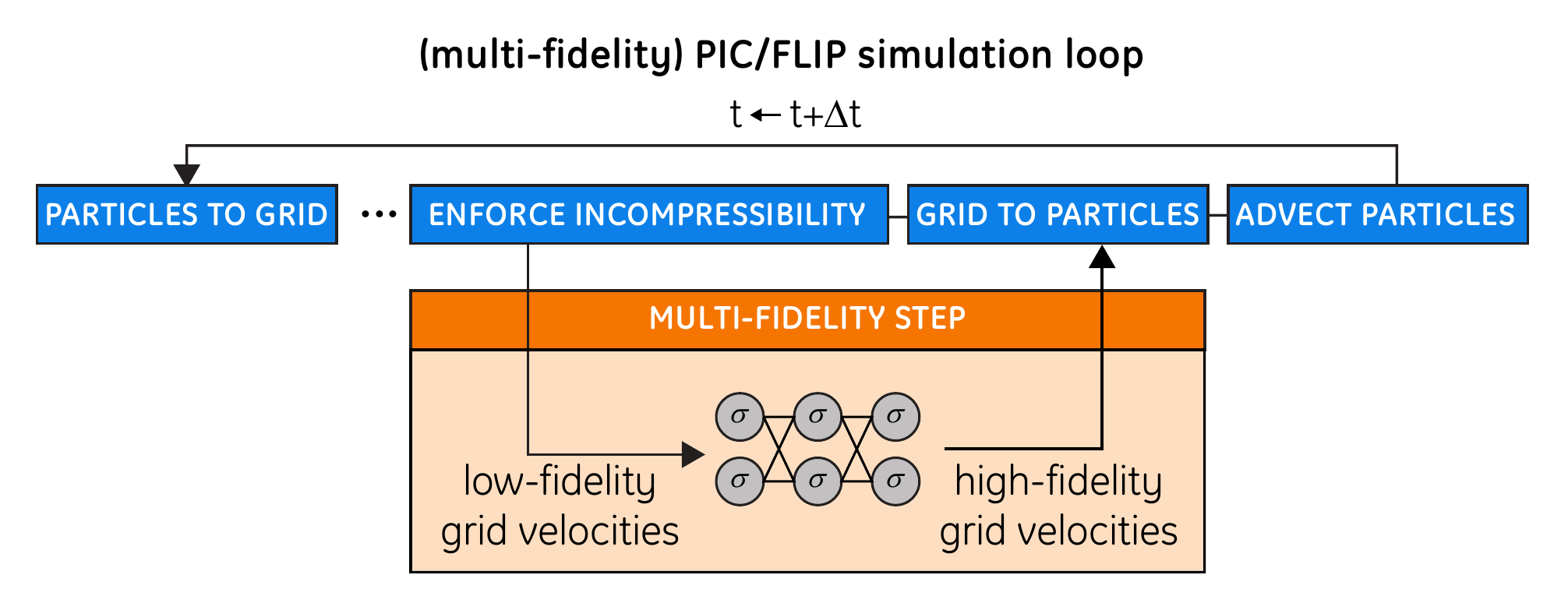}
\caption{\label{fig:PICFLIPoverview} Steps in the simulation loop for a single time step in the (multi-fidelity) PIC/FLIP solver.}
\end{figure}

Our approach is to incorporate a deep neural network in the PIC/FLIP solver that effectively increases the resolution of the computational grid by mapping a coarse-grid velocity field to its corresponding fine-grid counterpart, reducing the dominant errors that are caused by computing the incompressible velocity field and grid to particle transfer on a coarse grid. The resulting PIC/FLIP solver will be subsequently called the multi-fidelity PIC/FLIP, as it utilises both low and high-fidelity data to train the deep neural network. A schematic overview of the steps in a low-fidelity and multi-fidelity PIC/FLIP solver is shown in figure \ref{fig:PICFLIPoverview}.
\\
In order to train a neural network, we need to specify \cite{goodfellow_deep_2016, aggarwal_neural_2018}:
\begin{itemize}
\item Why this approach?
\item Define low-fidelity and high-fidelity,
\item Input/output quantities for the neural network,
\item Training data,
\item Neural network architecture,
\item Training procedure.
\end{itemize}
The steps are discussed individually in the next subsections.

\subsection{Why this approach?}
The core idea is to learn low-level features of the low-fidelity simulations and map these to the corresponding fine-grid counterpart at every time step. Recent work shows that neural networks are perfectly suited for predicting fluid flows \cite{pinn1, pinn2, pinn3}. However, they are often used non-intrusively as post-processing tools or as standalone solvers. Strictly enforcing physical laws inside a neural network is challenging and often not possible. Using neural networks intrusively in a solver though allows the neural network to optimally utilise physical laws. Enhancing coarse-grid solutions by using low-level features is not new. It is done in for instance turbulence simulations using Large-Eddy Simulation (LES) \cite{LES}, where subgrid features on a grid are modelled using quantities on that very only grid.

\subsection{Defining low and high-fidelity PIC/FLIP}
In order to apply our method for enhancing a low-fidelity PIC/FLIP solver, we need to define low and high-fidelity. As the computational bottleneck occurs in the computations on the grid, both the low and high-fidelity use the same number of particles, but differ in grid resolution. To clarify, a high-fidelity simulation corresponds to simulations on a fine-resolution computational grid with resolution $\Delta s_{HF}$ where we place $n_{p,\text{init}}=8$ particles in a wet cell during the initialisation, while a low-fidelity simulation corresponds to computations on a coarse-resolution computational grid with resolution $\Delta s_{LF}$ and where we place $8\left(\frac{\Delta s_{LF}}{\Delta s_{HF}}\right)^3$ particles in a wet cell during initialisation. Hence, the low-fidelity simulations have the same number of particles as the high-fidelity simulations and the combination of upscaling the resolution of the low-fidelity velocity grid and a large number of particles may result in an increase in accuracy. The grid resolutions $\Delta s_{HF}$ and $\Delta s_{LF}$ are specified in section 4. Lastly, the time step of the enhanced low-fidelity solver needs be such that the simulation remains stable when upscaling the velocity field. Therefore, we take the low-fidelity time-step the same as the high-fidelity time-step, which is chosen such that the high-fidelity simulation is stable. Even though a stable high-fidelity time-step is often orders of magnitude smaller than its low-fidelity counterpart, an upscaled low-fidelity solution is still beneficial as most of the computations are performed on a coarse grid which significantly reduces the time it takes to perform a single time-step when compared to high-fidelity computations.

\subsection{Input/output quantities for the neural network}
Our approach uses a deep neural network to enhance low-fidelity simulations at each time step. The inputs for the neural network have to be quantities that can be computed directly from the low-fidelity simulations at the current and/or previous time levels.

The choice of input and output quantities is motivated by the fact that the main error is caused by performing velocity calculations on a coarse grid. The input for the neural network is only the current state of the simulation as seen by the computational grid: 
\begin{itemize}
\item Scaled number of particles $P^s_{i,j,k}$ in each grid cell: 
\begin{equation}
P^s_{i,j,k}:=\min\left(1, \frac{P_{i,j,k}}{n_{p,\text{init}}\left(\frac{\Delta s_{LF}}{\Delta s_{HF}}\right)^3}\right)\ ,
\end{equation}
where $P_{i,j,k}$ is the number of particles in grid cell $i,j,k$.
\item The face velocities defined on the low-fidelity grid $\mbf{u}_{LF}$.
\end{itemize}
The scaled number of particles indicates if the cell is dry ($P^s_{i,j,k}=0$), fully wet ($P^s_{i,j,k}=1$), or partially wet ($0<P^s_{i,j,k}<1$). The $n_{p,\text{init}}$ in the definition of the scaled number of particles comes from the number of particles that are placed in a wet cell in the initialisation of a high-fidelity simulation (see section \ref{sec:init}), and the ratio $\frac{\Delta s_{LF}}{\Delta s_{HF}}$ accounts for the difference in the number of particles placed in a wet cell in the initialisation. The scaled number of particles is also bounded to be within the interval $[0, 1]$. One can use the positions of all particles directly as input for the neural network, but we opt for using the scaled number of particles per cell as input, which significantly reduces the dimensionality of the input when many particles are simulated, at a slight reduction in the information content. 

The outputs of the neural network are the new face velocities $\mbf{u}_{HF}$, which are defined on the high-fidelity grid, hence, effectively increasing the resolution of the grid. This new and improved velocity field is then fed back into the solver and used to calculate the new particle velocities in the \textit{grid to particles} transfer at the current time level with a smaller interpolation error \eqref{eq:interpolationerror}.

\subsection{Training data}
\label{sec:trainingdata}
The neural network is trained on a set of data that comprises both low- and high-fidelity data. The low-fidelity data is used as input for the neural network, while the high-fidelity data acts as a reference in the cost function, which is discussed in section \ref{sec:trainingprocedure}. The amount of data required in the training set may vary depending on the application and is often found heuristically \cite{goodfellow_deep_2016}. Furthermore, a validation set is constructed, which is used to tune the neural network architecture and hyperparameters, and a test set is used to test how well the network generalises.

\begin{figure}
\centering
\includegraphics[width = \textwidth]{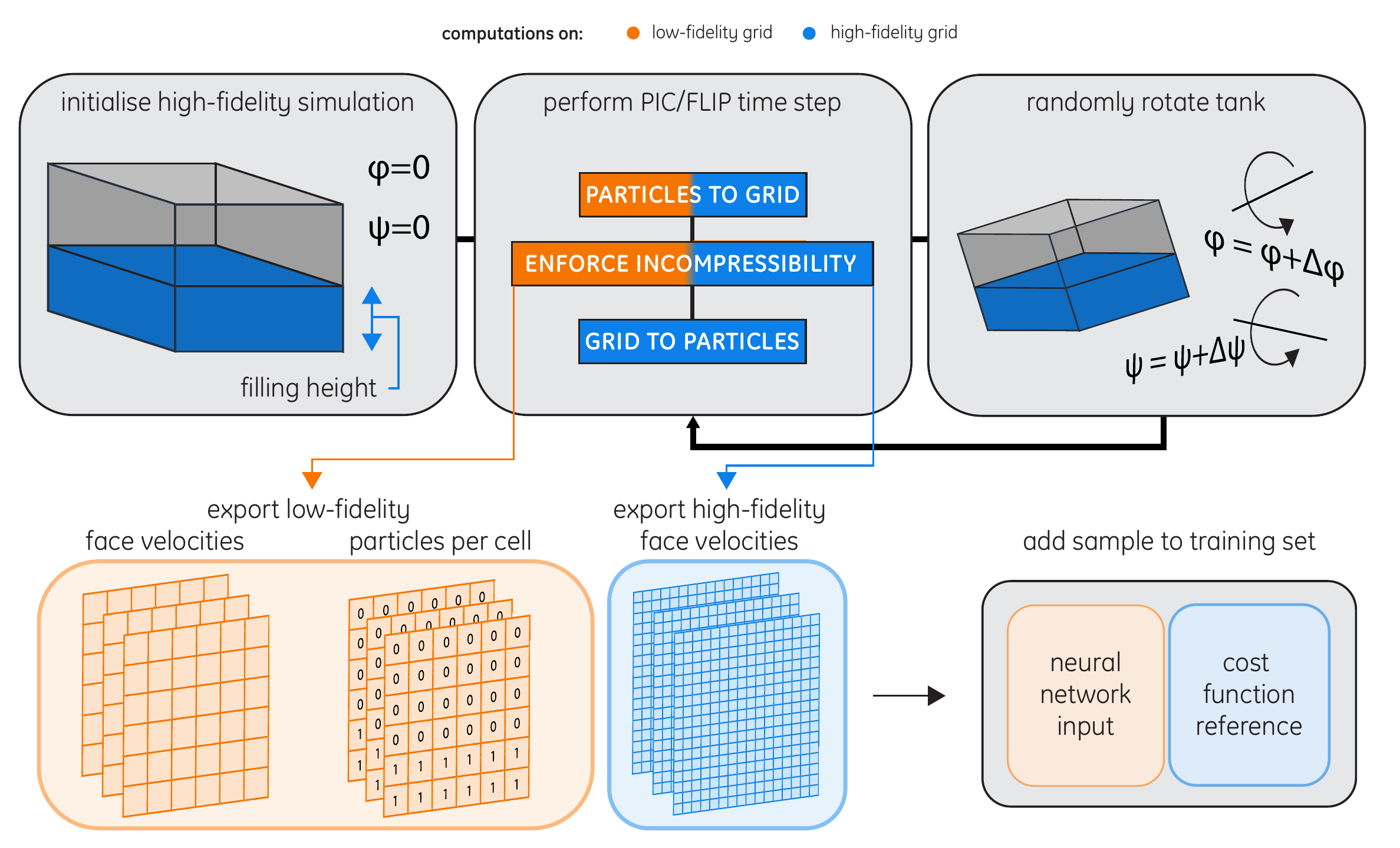}
\caption{\label{fig:dataSetConstruction} Schematic of how the data set is constructed that is used for training, validation, and testing. The magnitudes of the changes in angles $\Delta\psi$ and $\Delta\phi$ depend on the dimensions of the tank.}
\end{figure}

We focus on performing sloshing simulations \cite{ibrahim_liquid_2005}. Sloshing is the movement of a liquid contained inside a (possibly moving) object. These types of fluid flow simulations are relevant in many industry and CGI applications. For our application we construct the data set that is used for training, validation and testing, as is shown in figure \ref{fig:dataSetConstruction}.
\\
The procedure in figure \ref{fig:dataSetConstruction} performs a specified number of time steps and the motion of the rectangular tank is imposed by rotating the gravity vector in the simulation, i.e., $\psi(t=0)=\phi(t=0)=0$ and $\mbf{g}(t) = 10(\sin(\psi)\cos(\phi), \sin(\phi), -\cos(\psi)\cos(\phi))$, where the angles change randomly over time. After the data set has been constructed, we split the entire set in 70\% training, 20\% validation and 10\% testing samples \cite{goodfellow_deep_2016}.

The number of time steps, angle increments $\Delta \psi,\ \Delta\phi$, and the filling height determine how well the neural network performs and will be studied in detail in section \ref{sec:results}.

\subsection{Neural network architecture}
\label{sec:networkarchitecture}
Various options for the neural network type are available, e.g., fully-connected multilayer perceptron (MLP), convolutional neural network (CNN) or recurrent neural network (RNN), and determining which type to choose is often difficult. In general, the optimal neural network architecture depends on the application and a proper network type and architecture is often found by using expert knowledge.\\

\noindent\textit{\textbf{Fully-connected, convolutional, recurrent, or ...?}}\\
\noindent In this study, there is a clear spatial component in both the input and the output of the neural network while there is no temporal component in the randomly picked training data. As a result, we opt for using a CNN architecture due to its inherent spatial nature. Furthermore, it reduces the total number of degrees of freedom significantly as compared to using an MLP architecture. The spatial nature of CNNs is caused by using local filters that take a weighted sum of neighbouring velocity values to compute a new velocity value. If we increase the number of layers, then the amount of neighbouring velocity values, that are used to calculate a new velocity value, increases. We do not know beforehand how many neighbouring velocity values should be used for accurately approximating the new velocity values and the number of CNN layers is therefore considered a hyperparameter in our approach.\\

\begin{figure}
\centering
\includegraphics[width = \textwidth]{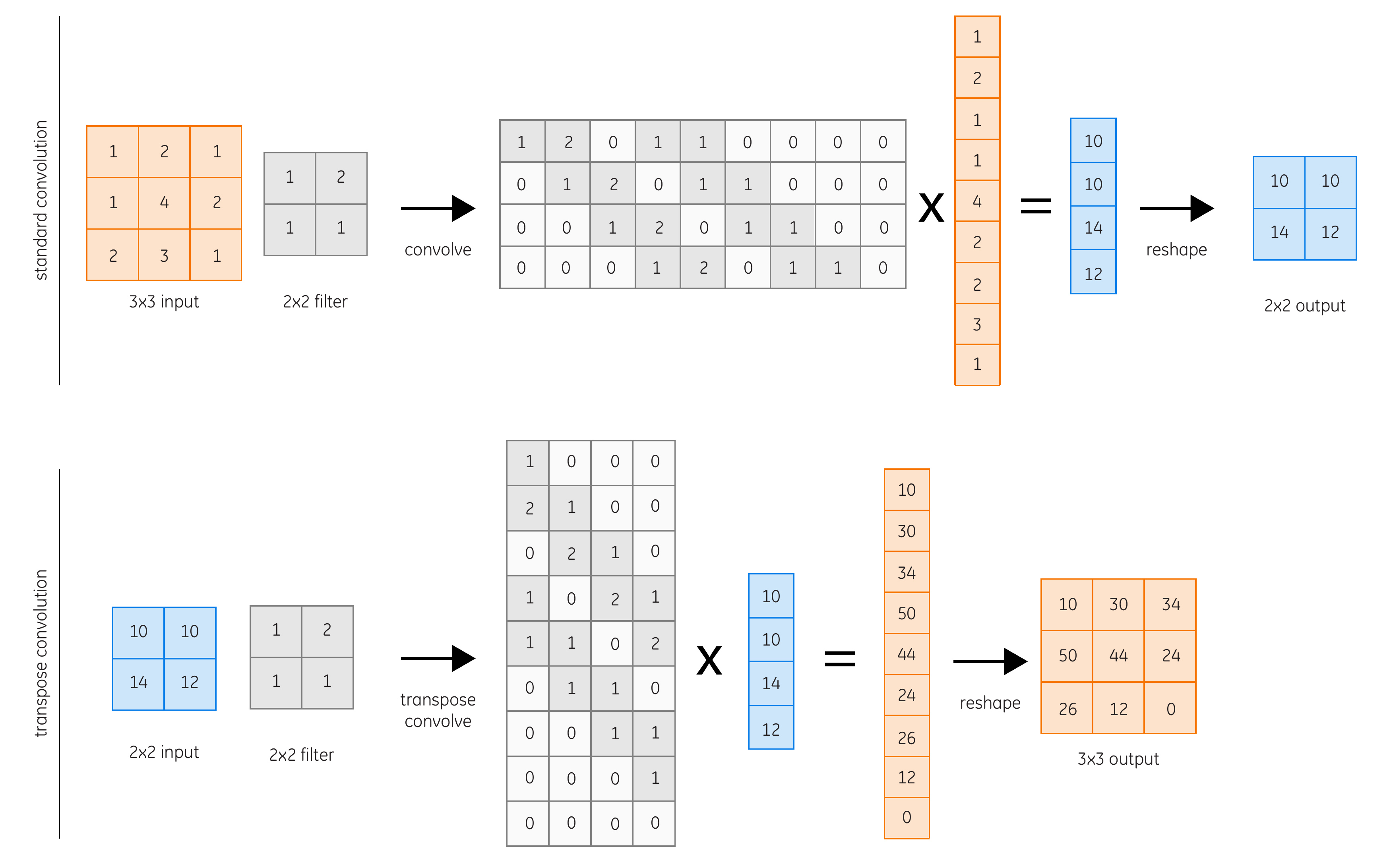}
\caption{\label{fig:convolutionexample} A standard 2D convolutional layer and a 2D transpose convolutional layer of a single kernel for a 1-channel input.}
\end{figure}

\noindent\textit{\textbf{How to increase the dimensions of the low-fidelity input to match the dimensions of the high-fidelity output?}}\\
\noindent When only using convolutional layers, the dimensionality of the input cannot be increased. This is problematic, because the goal of the neural network is to increase the effective grid resolution of the low-fidelity grid. To clarify, the low-fidelity input is of lower dimension than the high-fidelity output. This can be fixed by adding fully-connected layers after the convolutional part of the neural network. However these fully-connected layers significantly increase the number of degrees of freedom in the neural network and this leads to an increase of required training data. Constructing a larger training set requires more or longer high-fidelity simulations and is unwanted due to computational expense. The transposed convolutional layer, also known as deconvolutional layer \cite{deconv, deconv2}, solves this issue by performing a transposed matrix multiplication of the convolution matrix, which effectively increases the dimensions of the input and removes the need for adding fully-connected layers at the end of the network architecture. A schematic of the difference between a standard 2D convolutional layer and a 2D transposed convolutional layer is shown in figure \ref{fig:convolutionexample}.
\\
The transpose convolutional layer is characterised by the number of filters, the filter size and the stride\cite{goodfellow_deep_2016}.\\

\begin{figure}
\centering
\includegraphics[width = \textwidth]{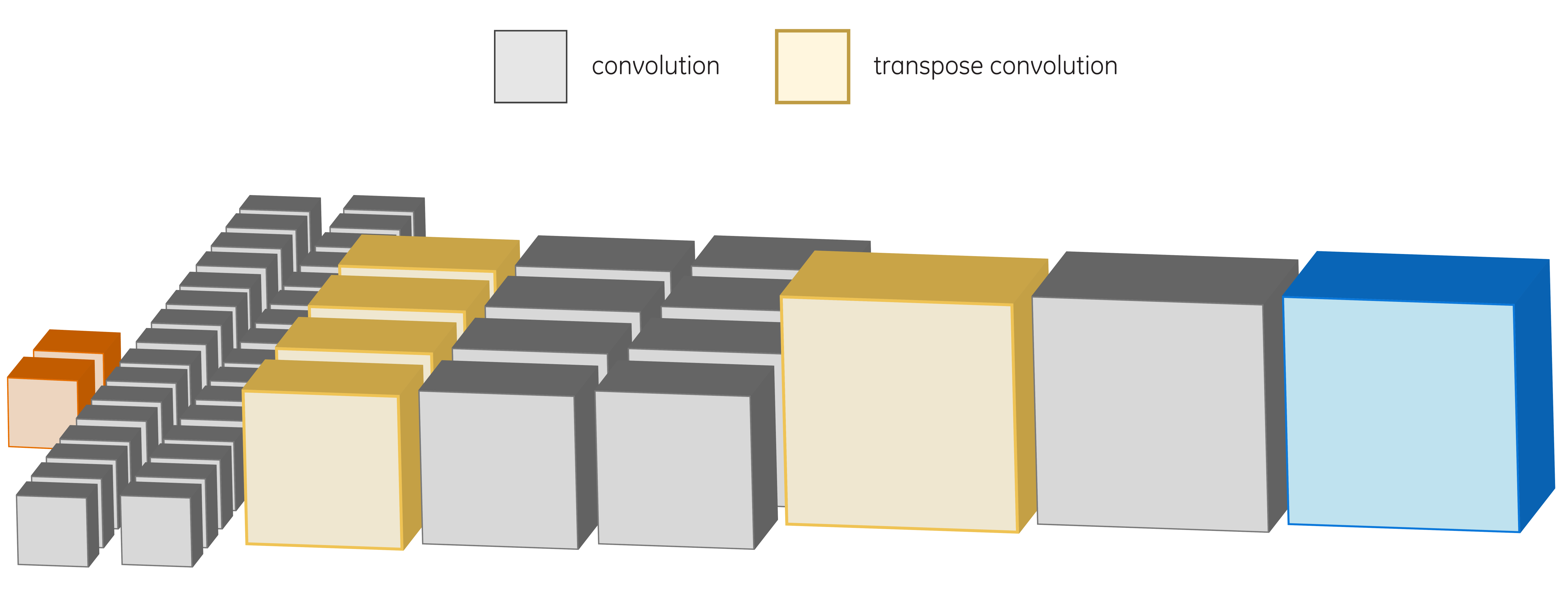}
\caption{\label{fig:completearchitecture} The complete neural network architecture with hyperparameter $N_{CNN}=2$. The architecture consists of a combination of convolutional and transpose convolutional layers.}
\end{figure}

\noindent\textit{\textbf{How to build the complete network architecture?}}\\
\noindent The complete neural network consists of a combination of both convolutional and transpose convolutional layers. The deconvolutional layers are used to increase the dimensionality of the neural network output, while the convolutional layers are used to effectively map the low-fidelity velocities to the high-fidelity velocities. A schematic overview of the complete network architecture is shown in figure \ref{fig:completearchitecture}.
\\
The architecture is defined as follows:
\begin{itemize}
\item The network starts with a number of 3D standard convolutional layers with a specific number of filters (discussed later).
\item After the first layer the network consists of groups of a single 3D transpose convolution, and $N_{CNN}$ 3D standard convolutional layers, where $N_{CNN}$ is a hyperparameter.
\begin{itemize}
\item The transpose convolutional layers increase the dimensions of their input with a factor 2 in every dimension, by using a $2\times 2$-filter and a stride 2.
\item The standard convolutional layers use a $3\times 3$-filter, which is a commonly used filter size in deep convolutional neural networks \cite{goodfellow_deep_2016}.
\item Zero padding is used at the standard convolutional layers to keep the input and the output dimensions of these layers unaltered.
\end{itemize}

\end{itemize}
The number of pairs is determined by the input (low-fidelity grid) and the output (high-fidelity grid) and the number of filters for each pair of deconvolution and convolutional layers decreases with a factor 4, ending with a single filter for the output layer representing the high-fidelity velocity field. For example, if the input corresponds to a low-fidelity grid of $16\times 16\times 16$ cells and the output corresponds to a high-fidelity grid of $256\times 256 \times 256$ cells, 4 pairs of transpose and standard convolutional layers are needed to match the dimensions of the output, where the first convolutional layer uses 64 filters. To assure that this makes sense, we require that the number of cells in each dimension on the high-fidelity grid is a multiple of 2 of the number of cells in the same dimension on the low-fidelity grid.

The network architecture is further characterised by the activation function used for each layer. In this work, we use a linear output activation to allow for unbounded values, and an exponential linear unit (ELU) \cite{clevert_fast_2015} activation function for the remaining layers:
\begin{equation}
\sigma(z) = \left\lbrace \begin{array}{l l}
z&,\ z>0\ ,\\
\gamma(e^z - 1)&,\ z\leq 0\ ,
\end{array}\right.
\label{eq:ELU}
\end{equation}
where $\gamma$ is a hyperparameter. We prefer the ELU activation function over conventional activation functions, such as ReLU and hyperbolic tangent, as it is easy to evaluate, suffers less from the dying neuron problem as compared to conventional ReLU, suffers less from the vanishing gradient problem as compared to hyperbolic tangent activation \cite{maas_rectifier_2013}, and because there is evidence that it speeds up training when compared to ReLU \cite{clevert_fast_2015}.

\subsection{Training the neural network}
\label{sec:trainingprocedure}
The cost function $c$ is a function of the weights and biases and indirectly depends on the hyperparameters. To prevent overfitting, regularisation will be used \cite{goodfellow_deep_2016}.

The randomly generated training data in section \ref{sec:trainingdata} is used to train neural networks with different choices for the hyperparameters discussed in section \ref{sec:networkarchitecture}. The cost function that is minimised is given by:
\begin{equation}
c_{\lambda}(W) = \sum_{i=1}^{N_t} \|(NN(\mbf{u}^{LF}_i, \mbf{p}_{\text{cell}, i}) -\mbf{u}^{HF}_i\|^2_2 + \lambda (\|W\|^2_2)\ ,
\end{equation}
where $W$ are the trainable parameters (filter coefficients, biases) of the neural network $NN$, and where $\lambda$ is the regularisation parameter, an additional hyperparameter for which a proper value needs to be determined during training. The cost function is minimised using the Adam optimizer with default parameter values $\beta_1=0.9,\ \beta_2=0.999$ and a step size $\alpha$ that is treated as a hyperparameter. Additionally, batch minimisation with batch size 256 is used to resolve out-of-memory errors, which may occur when passing a large dataset entirely for minimising the cost function. 

We find proper values for the hyperparameters $(N_{CNN}, \gamma, \lambda, \alpha)$ by performing a grid search on the tensor grid $H$ with 4 values for each hyperparameter. The 4 values for each hyperparameter are:
\begin{itemize}
\item \textbf{\textit{Number of CNN layers after a deconvolution layer}}: $N_{CNN}\in\{1, 2, 3, 4\}$ ,
\item \textbf{\textit{ELU shape parameter}}: $\gamma\in\{0.0001, 0.001, 0.01, 0.1\}$ ,
\item \textbf{\textit{Regularisation constant}}: $\lambda\in\{10^{-8}, 10^{-6}, 10^{-4}, 10^{-2}\}$ ,
\item \textbf{\textit{Optimiser step size}}: $\alpha\in\{0.0001, 0.001, 0.01, 0.1\}$ .
\end{itemize}
As a result, we need to train $4^4=256$ neural networks, one for each different combination of the hyperparameters. From this set of trained neural networks, the optimal network is defined as the one that has the smallest validation error
\begin{equation}
\arg\min_{(N_{CNN}, \gamma, \lambda, \alpha)\in H} \left[\sum_{i=1}^{N_v} \|(NN(\mbf{u}^{LF}_i, \mbf{p}_{\text{cell},i}) -\mbf{u}^{HF}_i\|^2_2\right]\ .
\label{eq:validationerror}
\end{equation}
The test set can be used to determine if the found optimal neural network generalises well to unseen data. The resulting neural network is used to enhance the low-fidelity PIC/FLIP simulations.

\section{Results}
\label{sec:results}
This section discusses the results of training and applying a multi-fidelity PIC/FLIP solver for fluid sloshing. The Tensorflow library for Python3 \cite{abadi_tensorflow:_2016} is used for constructing and training the neural networks, and it runs on 4 NVIDIA GTX 1080Ti GPUs. The PIC/FLIP solver is also implemented to run on 4 NVIDIA GTX 1080Ti GPUs. The section is divided into four parts:
\begin{itemize}
\item Enhancing low-fidelity PIC/FLIP sloshing simulations.
\item Generalisation capabilities for solver parameters.
\item Generalisation to a 3D dambreak problem.
\item A discussion of weaknesses of the multi-fidelity PIC/FLIP solver.
\end{itemize} 
For all simulations we take a computational domain $(x,y,z)\in[0, 10]\times[0, 5]\times[0, 5]$, unless stated otherwise. The high-fidelity PIC/FLIP solutions are computed with $\Delta s = 1/50$ ($500\times 250\times 250$ cells) and simulate up to 250 million particles. The low-fidelity PIC/FLIP solutions are computed with $\Delta s = 1/10$ ($100\times 50\times 50$ cells) and simulate up to 2 million particles. The grid resolution of the low-fidelity simulations is chosen such that it is able to run in real-time, while the high-fidelity simulation grid contains 125 times more grid cells and can not be performed in real-time. The time step needs to be the same for both fidelities and is set to $\Delta t=1/350$ which results in stable low- and high-fidelity simulations. Lastly, the PICness $f$ is set to 0.99 for both fidelities, which is a commonly used value in literature for simulating free surface water flow \cite{zhu_animating_2005}. 

\subsection{Enhancing low-fidelity PIC/FLIP sloshing simulations}
\label{sec:resultsSloshing}
We show that the proposed multi-fidelity approach can indeed enhance low-fidelity PIC/FLIP sloshing simulations.\\

\noindent\textit{\textbf{The type of solutions that are enhanced}}\\
In order to study the effectiveness of a multi-fidelity PIC/FLIP solver, we prescribe a tank motion with $\mbf{g} = 9.81(\sin(\psi)\cos(\phi),\sin(\phi), -\cos(\psi)\cos(\phi))$, where $\psi$ and $\phi$ are given by:
\begin{equation}
\psi(t) = \left\lbrace \begin{array}{l l}
\sin(2t)&,\ t<2\pi\ ,\\
0&,\ t\geq 2\pi\ ,
\end{array}\right.\ \ \phi(t) = \left\lbrace \begin{array}{l l}
\sin(t)&,\ t<2\pi\ .\\
0&,\ t\geq 2\pi\ .
\end{array}\right.
\label{eq:motion}
\end{equation}
We choose a deterministic motion as we have full control over the sloshing phenomenon encountered in the simulation, which is not possible when picking a random tank motion. This particular motion causes heavy liquid sloshing in the tank and is perfectly suited for testing our method. The filling height is varied to study generalisation capabilities of the method, and ranges between $(0, 100)\%$.

\noindent\textit{\textbf{Construction of the training set}}\\
The procedure for constructing the training data is shown in figure \ref{fig:dataSetConstruction}. A summary of all training parameters is given by:
\begin{itemize}
\item Data set: $10^6$ generated training samples (see figure \ref{fig:dataSetConstruction}) with a filling height of 40\%, $\Delta\psi,\Delta\phi$ are picked randomly every time step in the ranges $[-2\Delta t, 2 \Delta t]$, $[-\Delta t, \Delta t]$, respectively.
\begin{itemize}
\item 70\% training,
\item 20\% validation,
\item 10\% testing,
\end{itemize}
\item Training parameters: batch-size=256 , number of epochs=1000.
\end{itemize}
The training set only comprises samples with a single filling height of $40\%$, which corresponds to a filling height that allows for heavy sloshing motions and therefore a wide range of fluid configurations. This allows us to study if the neural network generalises well to cases with a different filling height. The range for $(\Delta\psi,\Delta\phi)$ corresponds to the same oscillation frequency as the one in \eqref{eq:motion}. However, the probability that the tank motion \eqref{eq:motion} is comprised in the training data is negligible.

\begin{figure}[hbt!]
\includegraphics[width = \linewidth]{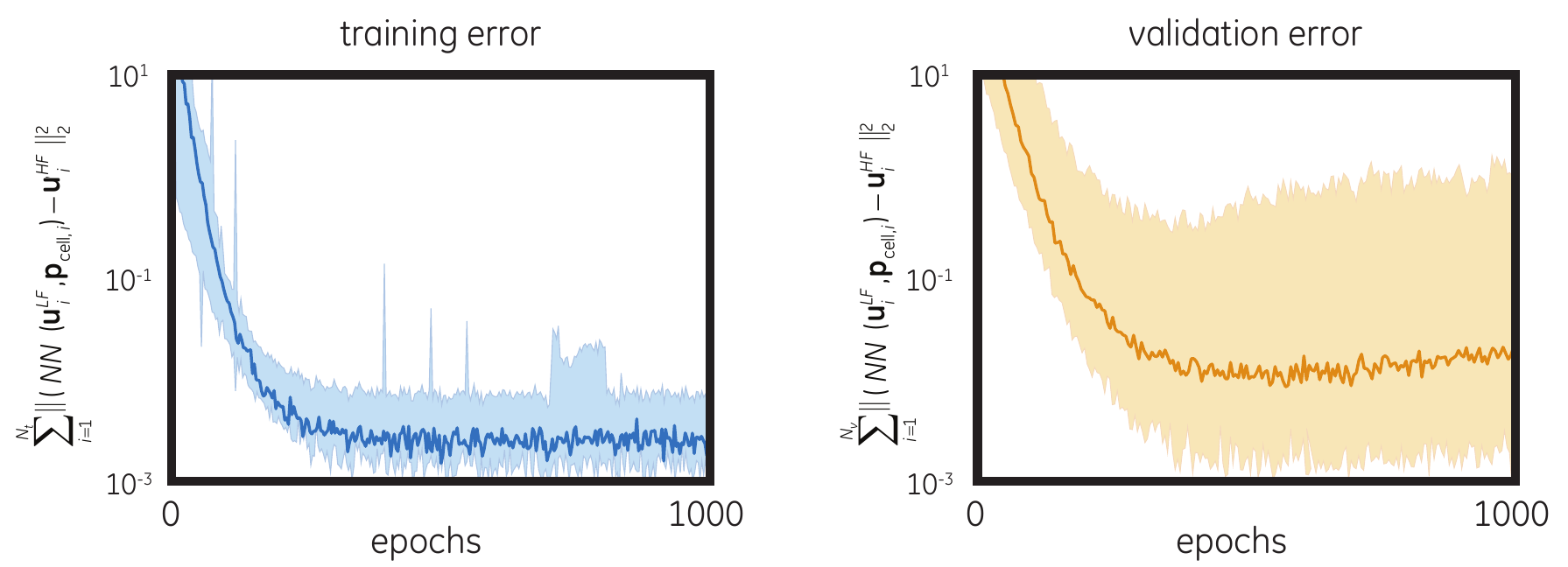}
\caption{\label{fig:trainingerror} The mean and envelope of both the training and validation error. The envelope shows the minimum and maximum values over all 256 trained neural networks.}
\end{figure}

\noindent\textit{\textbf{Training the neural networks}}\\
A total of 256 neural networks need to be trained (1 for each set of hyperparameters), and the resulting training and validation errors are shown in figure \ref{fig:trainingerror}.
\\
We clearly see that all 256 trained neural networks converge to a similar error in the training set, while the variation in the validation error is more pronounced. This indicates under/overfitting and a lack of generalisation capabilities for specific hyperparameter values. Remarkable is that the 103 networks with the largest validation errors are the networks where $N_{CNN}=1,2$ which indicates that these relatively local multi-fidelity velocity approximations are not capable of generalisation outside the training set. This indicates that to approximate a local multi-fidelity velocity, we may need a large area of surrounding low-fidelity velocities. This global character of the required approximation may be caused by the incompressibility of the fluid, which leads to a coupling of velocities by means of the pressure Poisson equation \eqref{eq:pressureequation} which is also felt globally. The hyperparameters $(N_{CNN}, \gamma, \lambda, \alpha) = (3, 0.01, 10^{-4}, 0.001)$ satisfy \eqref{eq:validationerror}, and the trained neural network with these hyperparameter values is used in the remainder of this paper.

\begin{figure}[hbt!]
\includegraphics[width = \linewidth]{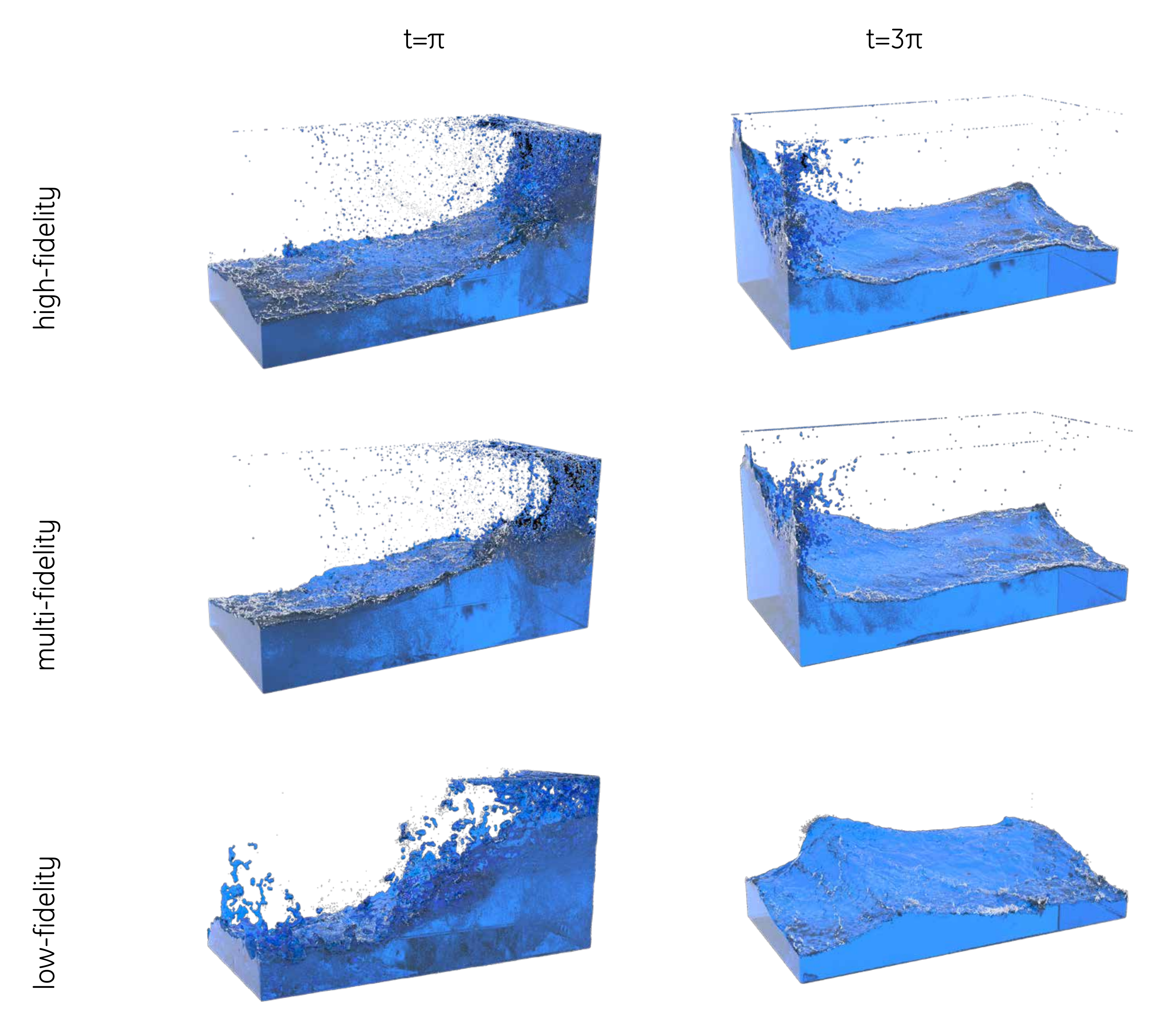}
\caption{\label{fig:results1} Rendered fluid surface for the three different fidelities at $t=\pi, 3\pi$ with a filling height of 40\%. The white particles are passive tracers that represent foam and spray \cite{foam} and which are added in a post-processing stage by detecting where air entrapment is occurring during the simulation, i.e., parts with high velocity gradients and areas where the fluid surface is concave.}
\end{figure}

\noindent\textit{\textbf{Enhancing solution with the same filling height as used during training}}\\
First, we start by using the trained neural network to enhance a sloshing simulation with the motion given by \eqref{eq:motion} given a filling height of 40\%. Notice that this is the same filling height that is used for constructing the training set and is therefore considered a consistency test for the trained neural network. The low-fidelity, multi-fidelity, and high-fidelity PIC/FLIP simulations are compared at $t=\pi,3\pi$. The rendered comparison is shown in figure \ref{fig:results1}.
\\
When we compare both low- and multi-fidelity solution to the reference high-fidelity solution, we see a clear qualitative improvement when using the multi-fidelity solver. Notice that the multi-fidelity solution at $t=\pi$ shows smaller scale droplets when compared to the low-fidelity solution, which is caused by effectively increasing the resolution of the grid. Furthermore, the multi-fidelity and the high-fidelity solutions at $t=3\pi$ show a similar fluid surface, while the low-fidelity fluid surface has a completely different shape.

\begin{figure}[hbt!]
\centering
\includegraphics[width = \linewidth]{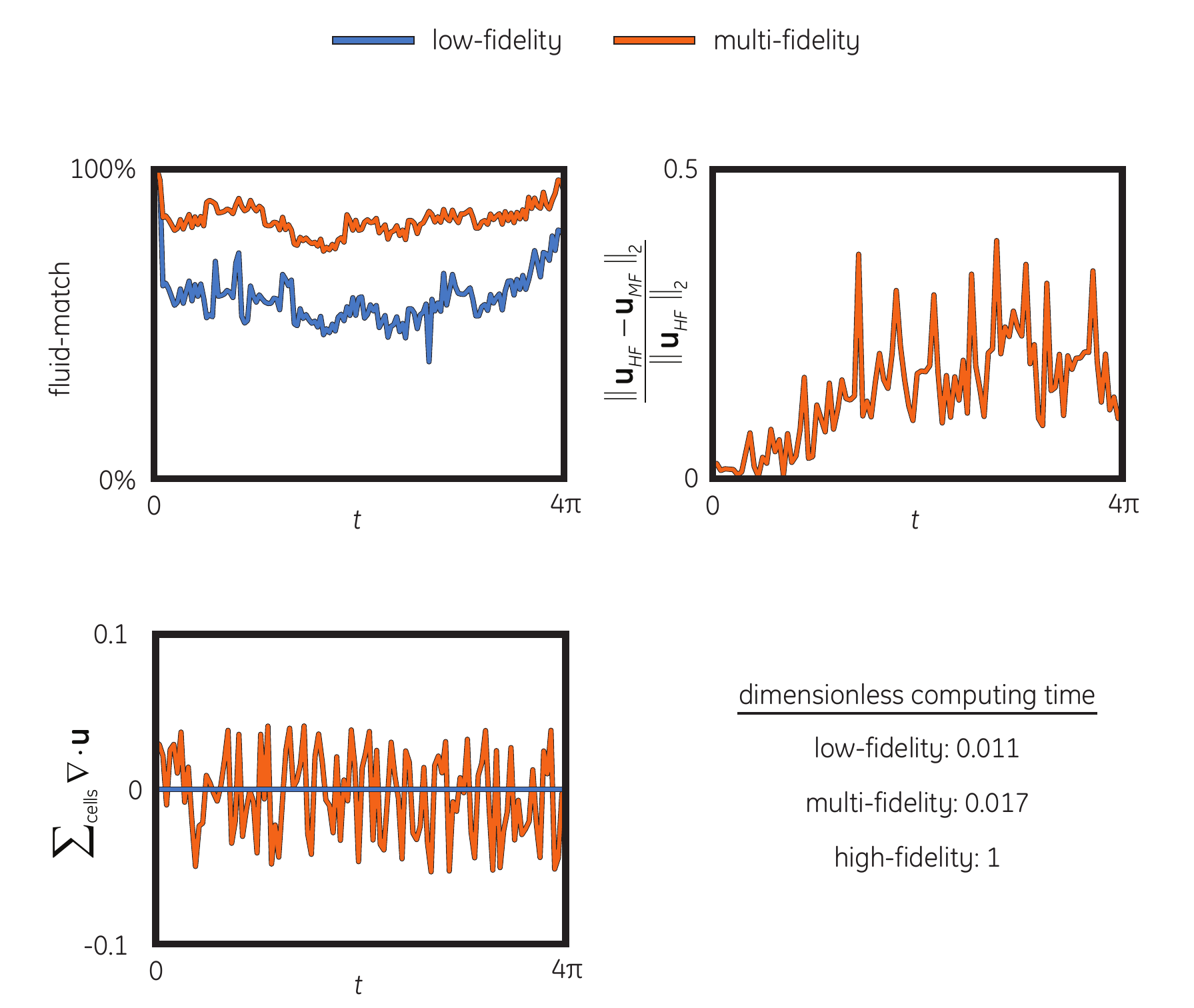}
\caption{\label{fig:results2} Fluid-match, multi-fidelity incompressibility computed with \eqref{eqPICFLIP:discreteincompressible}, and multi-fidelity (MF) relative velocity error.}
\end{figure}

To study the difference between the three fidelities quantitatively, we perform a simulation until $t=4\pi$ for each of the three fidelities, and compare the so-called fluid-match. The fluid-match is defined as the percentage of the total volume that coincides with the high-fidelity solution, resulting in a fluid-match that is close to 100\% if a solution is close to the high-fidelity reference and almost 0\% when the simulation is completely off. This is used as a measure to quantify how close a solution is to the reference high-fidelity solution, i.e., if the fluid-match is $100\%$, then the fluid surface in the performed simulation is the same as the high-fidelity fluid surface. Notice that this does not necessarily mean that the neural network velocities are the same as the high-fidelity velocities. A quantitative comparison between the different fidelities is shown in figure \ref{fig:results2}.
\\
A smooth initial part of the fluid-match resembles the period before the fluid impacts the boundaries of the domain. After the first impact, the fluid-match oscillates. When simulating for a longer period, the fluid-match converges back to almost $100\%$, due to the fluid coming to rest again. Notice that the fluid-match is not exactly 100\% after the fluid being almost at rest, which is due to the fact that the PIC/FLIP solver does not preserve fluid volume in the presence of large particle velocities. These large velocities may result in particles being more densely clustered which causes the fluid volume to shrink. We clearly see a significant improvement in fluid-match when using the multi-fidelity PIC/FLIP solver, while not significantly increasing computational time. The neural network produces slightly compressible face-velocities, which may also cause the shrinking and growing of the total fluid volume. Lastly, the multi-fidelity and high-fidelity velocity fields show a gradual increase of mismatch in velocity values. The oscillating behaviour with large outliers of the velocity mismatch is caused by the difference in droplet distributions, as shown in figure \ref{fig:results1} at $t=\pi$. These droplets often have high velocities and a slight difference in droplet positions may cause a large mismatch in the face velocities. However, because these large outliers in velocity mismatch are caused by droplets, this large mismatch is not seen in the fluid-match, as they only affect the flow locally in a small part of the computational domain. \textbf{\textit{To summarise, the multi-fidelity solver significantly increases accuracy in terms of fluid-match for a sloshing simulation with the same filling height that was used during training, but generates slightly compressible velocity fields.}}

\begin{figure}[hbt!]
\centering
\includegraphics[width = \linewidth]{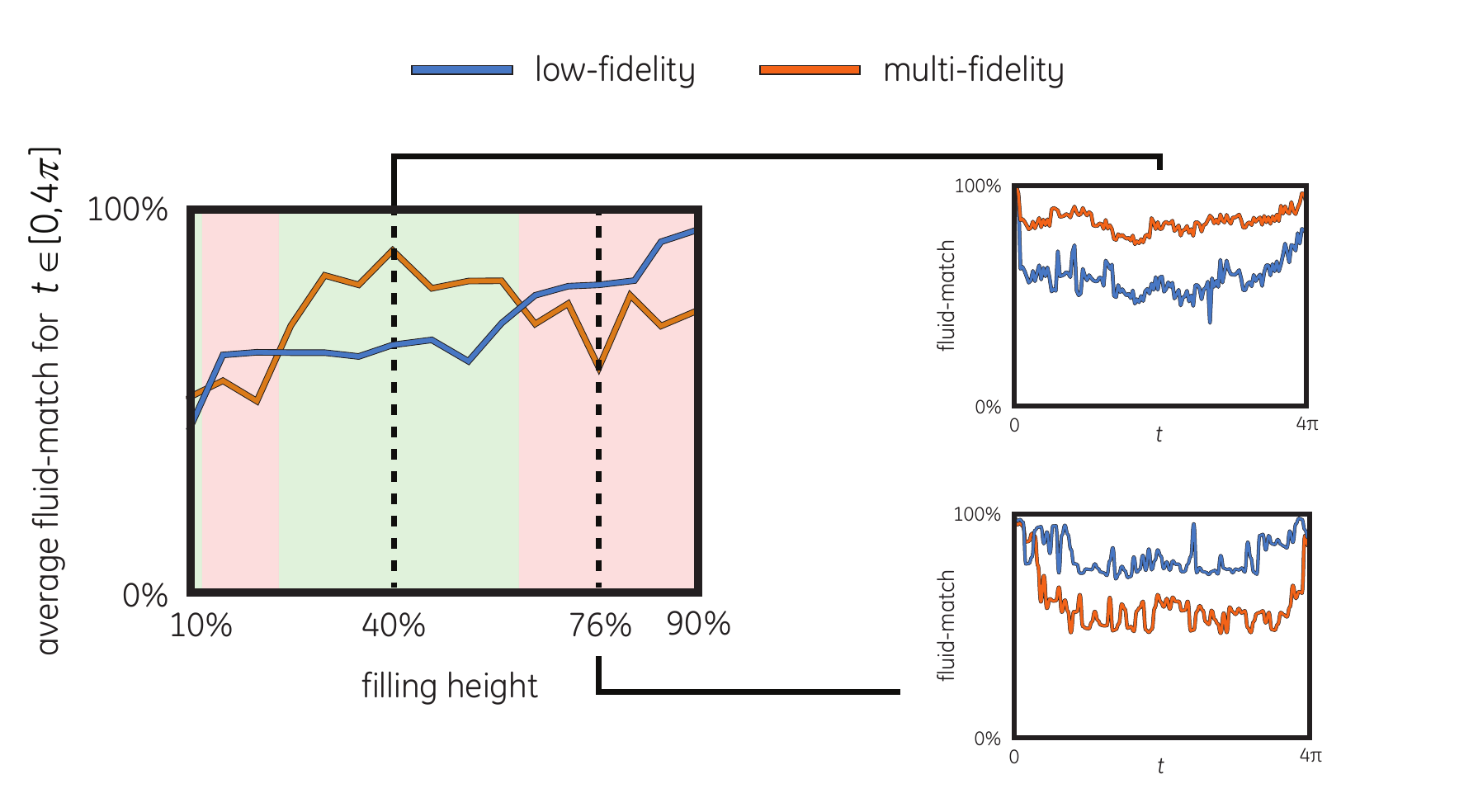}
\caption{\label{fig:results3} Error in enhanced sloshing simulations for filling heights that were not used during training.}
\end{figure}

\noindent\textit{\textbf{Enhancing solutions with different filling heights}}\\
Enhancing solutions where the filling height differs from the filling height that was used during training is more difficult, as the trained neural network has never seen inputs that are associated with different filling heights. Figure \ref{fig:results3} shows how well the neural network generalises for filling heights outside the training set.
\\
The multi-fidelity solver outperforms the low-fidelity solver for a significant portion of filling heights (indicated by the green area in figure \ref{fig:results3}). The low-fidelity fluid-match shows improvement in fluid-match with increasing filling height, which eventually leads to out-performance of the multi-fidelity solver. This increasing trend is caused by the fact that fluid sloshing affects the fluid-match less and less when the filling height goes to 100\% (completely filled tank without fluid sloshing for which low-fidelity is already close to high-fidelity simulation). 

\textbf{\textit{To summarise, the multi-fidelity solver is able to the generalise to other filling heights, but leads to inaccurate results when deviating too much from the training set. Furthermore, without a high-fidelity reference, which is not available after training, it is difficult to predict the limits of the generalisation capabilities of the trained neural network. This is a common problem in machine learning and is an active field of research.}}

\subsection{Generalisation capabilities for solver parameter changes}
\label{sec:generalise}
\begin{figure}[hbt!]
\centering
\includegraphics[width = \linewidth]{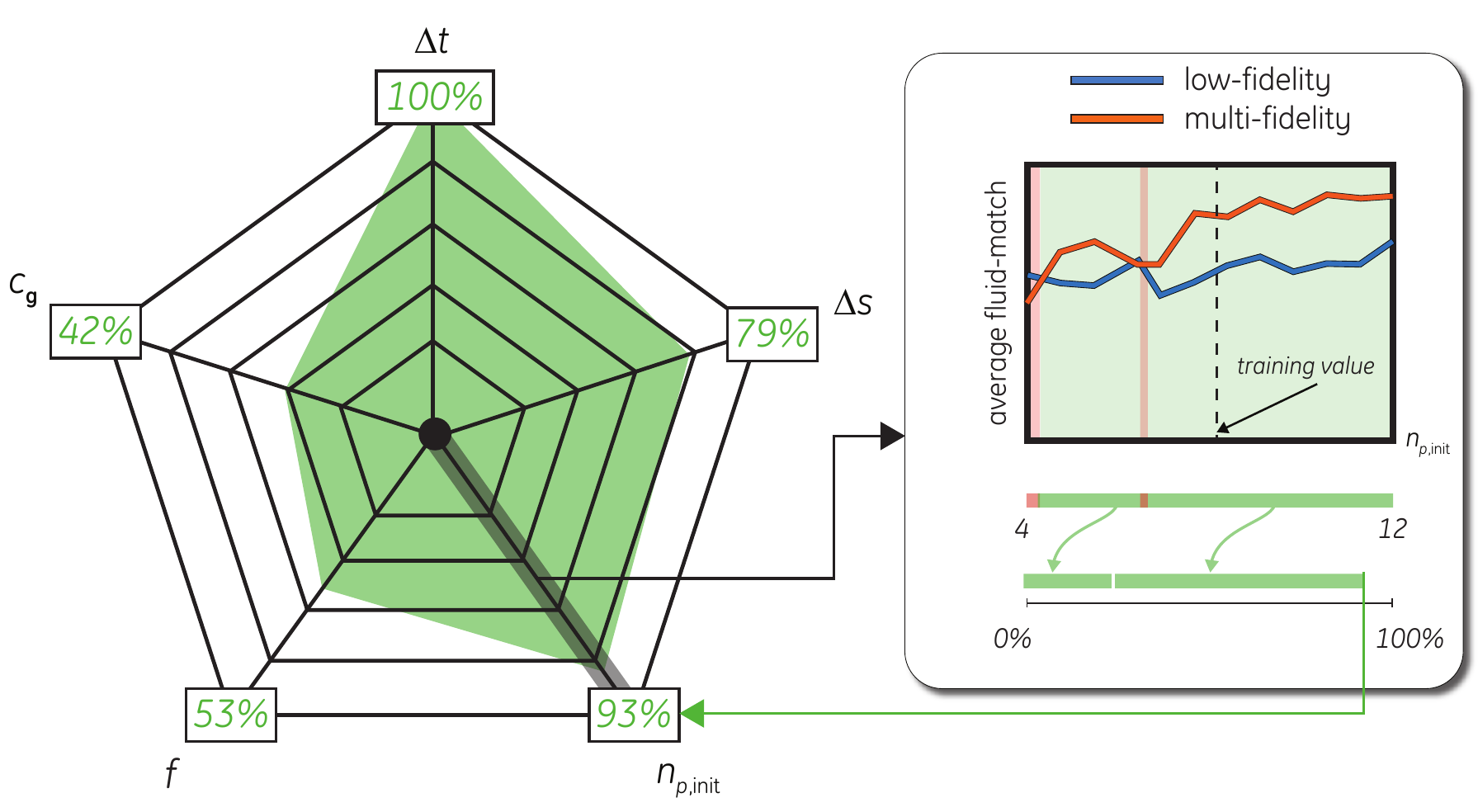}
\caption{\label{fig:results4} Generalisation capabilities for different solver parameters. The green regions indicate the fraction of parameter values for which the multi-fidelity fluid-match is higher than the low-fidelity fluid-match for the sloshing simulation for $t\in[0, 4\pi]$.}
\end{figure}

The filling height is only one out of several parameters that were used in the sloshing test case. Studying how the multi-fidelity solver generalises when changing solver parameters provides information on how sensitive the neural network is to the training data. In this section we show how the multi-fidelity solver generalises when changing the following parameters:
\begin{itemize}
\item \textbf{\textit{Time step}}: $\Delta t \in [\frac{1}{275}, \frac{1}{500}]$.
\begin{itemize}
\item The time step is picked such that the high-fidelity simulations are stable, which makes it easy to compare all three fidelities.
\end{itemize}
\item \textbf{\textit{Computational domain size scaling}}: $c_{\Delta s}\in[\frac{1}{2}, \frac{3}{2}]$. 
\begin{itemize}
\item While keeping the number of cells in the low- and high-fidelity simulations the same, we can change $\Delta s\rightarrow c_{\Delta s} \Delta s$ to effectively increase the size of the computational domain.
\end{itemize}
\item \textbf{\textit{Number of particles in wet cell during initialisation}}: $n_{p, \text{init}}\in[4, 12]$.
\begin{itemize}
\item The wet cells in the multi-fidelity simulation are initialised with $n_{p, \text{init}}\left(\frac{\Delta s_{LF}}{\Delta s_{HF}}\right)^3$ particles ($n_{p, \text{init}}=8$ was used for training).
\end{itemize}
\item \textbf{\textit{PICness parameter}}: $f\in[0, 1]$.
\begin{itemize}
\item Determines how diffusive the fluid motions are. Changing this value significantly changes the behaviour of the fluid ($f=0.99$ was used for training).
\end{itemize}
\item \textbf{\textit{Gravity vector scaling}}: $c_{\mbf{g}}\in[5, 15]$.
\begin{itemize}
\item The gravity vector will be defined as $\mbf{g}=c_{\mbf{g}}(\sin(\psi)\cos(\phi),\sin(\phi), -\cos(\psi)\cos(\phi))$ ($c_{\mbf{g}}=9.81$ was used for training). The value for $c_{\mbf{g}}$ determines the strength of the gravitational force and changing this value significantly changes fluid behaviour.
\end{itemize}
\end{itemize}
The fluid-matches for the low- and multi-fidelity simulations are compared for the sloshing simulation characterised by the sloshing motion \eqref{eq:motion} for $t\in[0, 4\pi]$. The results are shown in figure \ref{fig:results4}.
\\
We clearly see that the multi-fidelity solver generalises well for all solver parameters apart from the PICness parameter $f$ and gravity scaling constant $c_{\mbf{g}}$. The parameters $\Delta t$, $\Delta s$ and $n_{p, \text{init}}$ do not significantly change the fluid motions and the neural network will receive inputs that are similar to the training data. As a result, the multi-fidelity solver attains a significant increase in accuracy when compared to the low-fidelity solver. The parameters $f$ and $c_{\mbf{g}}$ do significantly change the fluid behaviour though, which results in neural network inputs that significantly deviate from the training data. 

\textbf{\textit{To summarise, for a large portion of the parameter value ranges the multi-fidelity solutions show an increase in accuracy when compared to the low-fidelity solutions. However, it is not possible to generalise this to cases where $c_g$ and $f$ deviate significantly from the training data.}}

\subsection{Generalisation to a 3D dambreak problem}
It was shown in sections \ref{sec:resultsSloshing} that the trained neural network, that was used in the multi-fidelity solver, was able to generalise to different filling heights and solver parameters values. However, the simulations were still similar in the sense that the prescribed motion \eqref{eq:motion} was the same and the initial fluid configuration was similar. In this section we show that the trained neural network is able to generalise to a different test case as well.

\begin{figure}[hbt!]
\centering
\includegraphics[width = \linewidth]{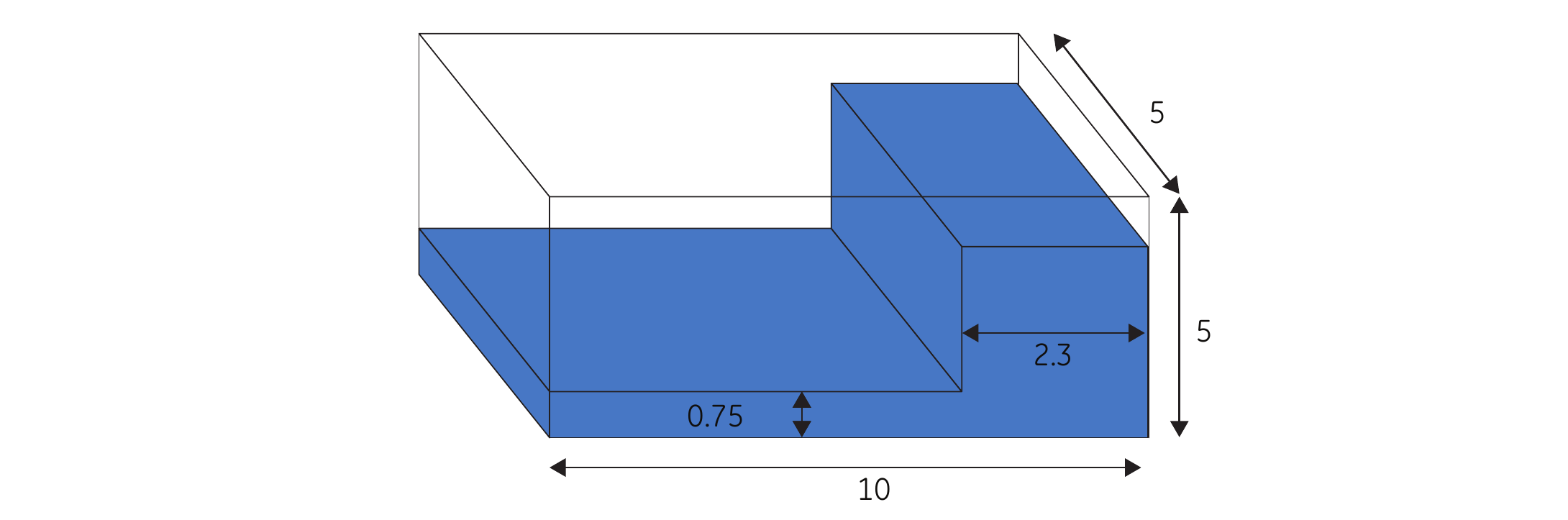}
\caption{\label{fig:dambreakinit} The initial fluid configuration of the wet dambreak problem.}
\end{figure}

\begin{figure}[hbt!]
\centering
\includegraphics[width = \linewidth]{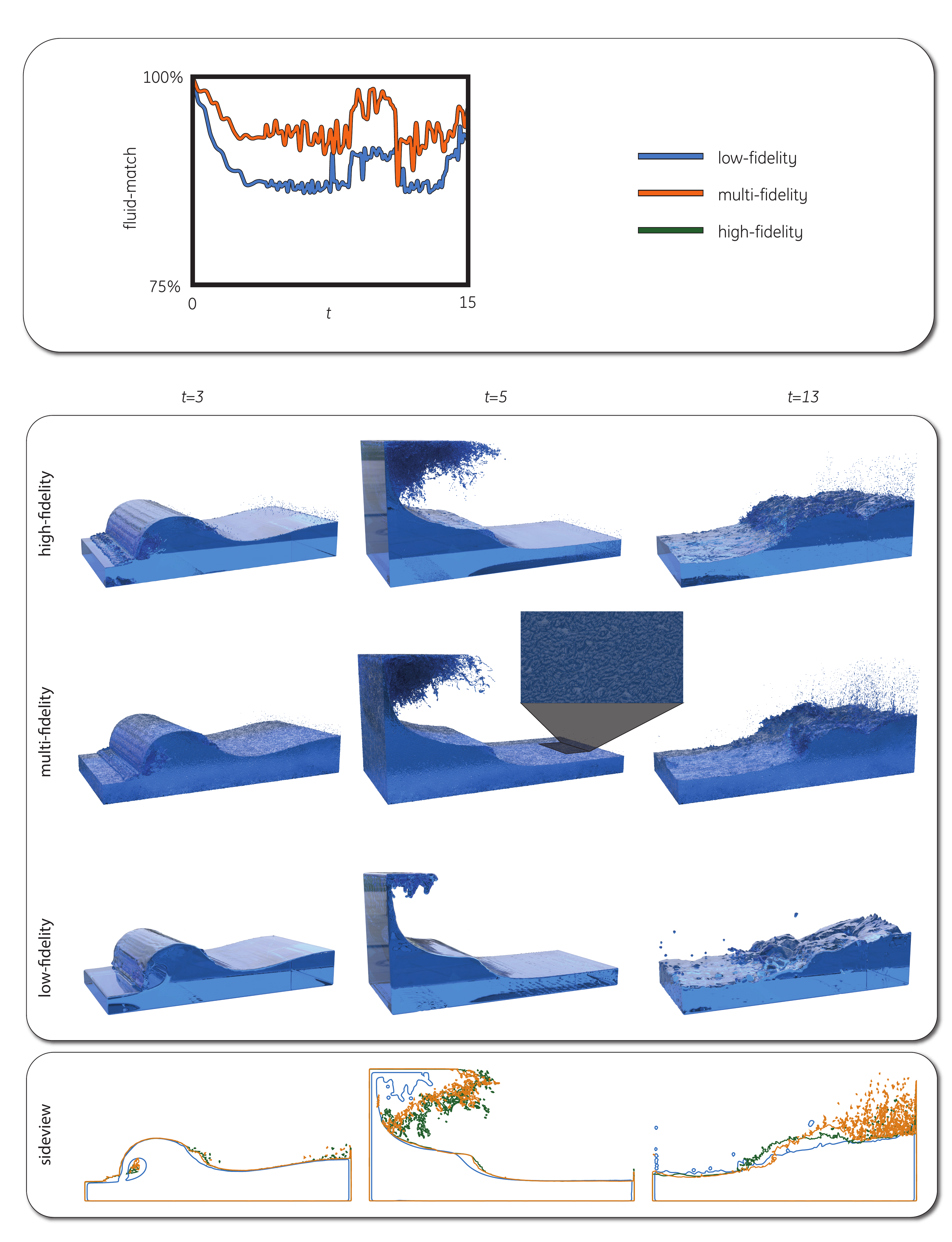}
\caption{\label{fig:dambreak_results1} A comparison of the low-, multi-, and high-fidelity solutions for the dambreak problem. The multi-fidelity solution at $t=5$ shows a close-up of the tiny oscillations on the fluid surface.}
\end{figure}

We consider a wet dambreak problem of which the initial fluid configuration is shown in figure \ref{fig:dambreakinit}.
\\
This particular configuration is chosen such that the total volume corresponds to a 30\% filling height. This allows us to study if the neural network generalises to a case with a filling height different from the one that was used for training, and was also initialised in a different configuration. The dambreak problem leads to a breaking wave that impacts the left domain boundary. The low and multi-fidelity solutions are compared for $t\in[0, 15]$, i.e., the period before, during, and after the wave impact. A qualitative and quantitative comparison of the three fidelities is shown in figure \ref{fig:dambreak_results1}.
\\
Both the low- and multi-fidelity simulation show good agreement with the high-fidelity solution. The effectiveness of the neural network approach is again emphasised by the fluid-match, which shows an increase in accuracy for the multi-fidelity solution, even for this dambreak problem for which the training set was not tailored. However, the multi-fidelity solution shows tiny oscillations on the fluid surface, which is caused by oscillatory velocity predictions. The frequency of the oscillation corresponds to the largest frequency that can be resolved on the high-fidelity computational grid . They immediately appear in the first few time steps of the multi-fidelity simulation and grow slightly until the wave impacts the domain boundary. This phenomenon is most likely caused by the configuration of the initial condition which is not comprised in the training set and therefore not seen by the neural network during training. Even though these oscillations are unwanted, they do not affect the accuracy of the multi-fidelity solution severely. In order to possibly remove the unwanted oscillations, one may enlarge the training set to encapsulate more test cases. 

\textbf{\textit{To summarise, these results indicate that the generalisation to other test cases is possible, but a carefully constructed training set may be needed to remove unwanted artefacts in the enhanced velocity field.}}

\subsection{Weaknesses of the multi-fidelity PIC/FLIP solver}
It was shown that our approach is able to generalise to solver parameters and test cases slightly outside the training set. However, the problem of generalisation to cases further away from the training data still exists. In this section we give an overview of the problems we encountered during training of the neural network and of the limitations of the proposed approach. In short the limitations of our multi-fidelity approach concern:
\begin{itemize}
\item Sensitivity to solver parameters.
\item Does not generalise to other test cases.
\item Difficult to obtain physical interpretation of the multi-fidelity approach.
\end{itemize}

\noindent\textit{\textbf{Sensitivity to solver parameters}}\\
As is shown in figure \ref{fig:results4}, the neural network is able to simulate cases with slightly different solver parameters. It is possible to construct a training set that comprises simulations with multiple solver parameters, which increases the range of solver parameters for which the multi-fidelity solver produces satisfactory results. However, as the training set is constructed with low- and high-fidelity data, we need to make sure that the high-fidelity simulations are stable, which results in a small time step. We chose to use this small time step for the low-fidelity simulations, which results in a multi-fidelity solver that also uses a similar small time step. An alternative is to construct the training set with low-fidelity simulations, computed with a large time step, and a high-fidelity counterpart, computed with a small time step. The resulting training set then comprises all low-fidelity data and the sub sampled high-fidelity simulation data at the same time levels as the low-fidelity simulation. This approach leads to unstable multi-fidelity solutions which indicates that the multi-fidelity solver needs to satisfy similar stability conditions as the high-fidelity solver.

\begin{figure}[hbt!]
\centering
\includegraphics[width = \linewidth]{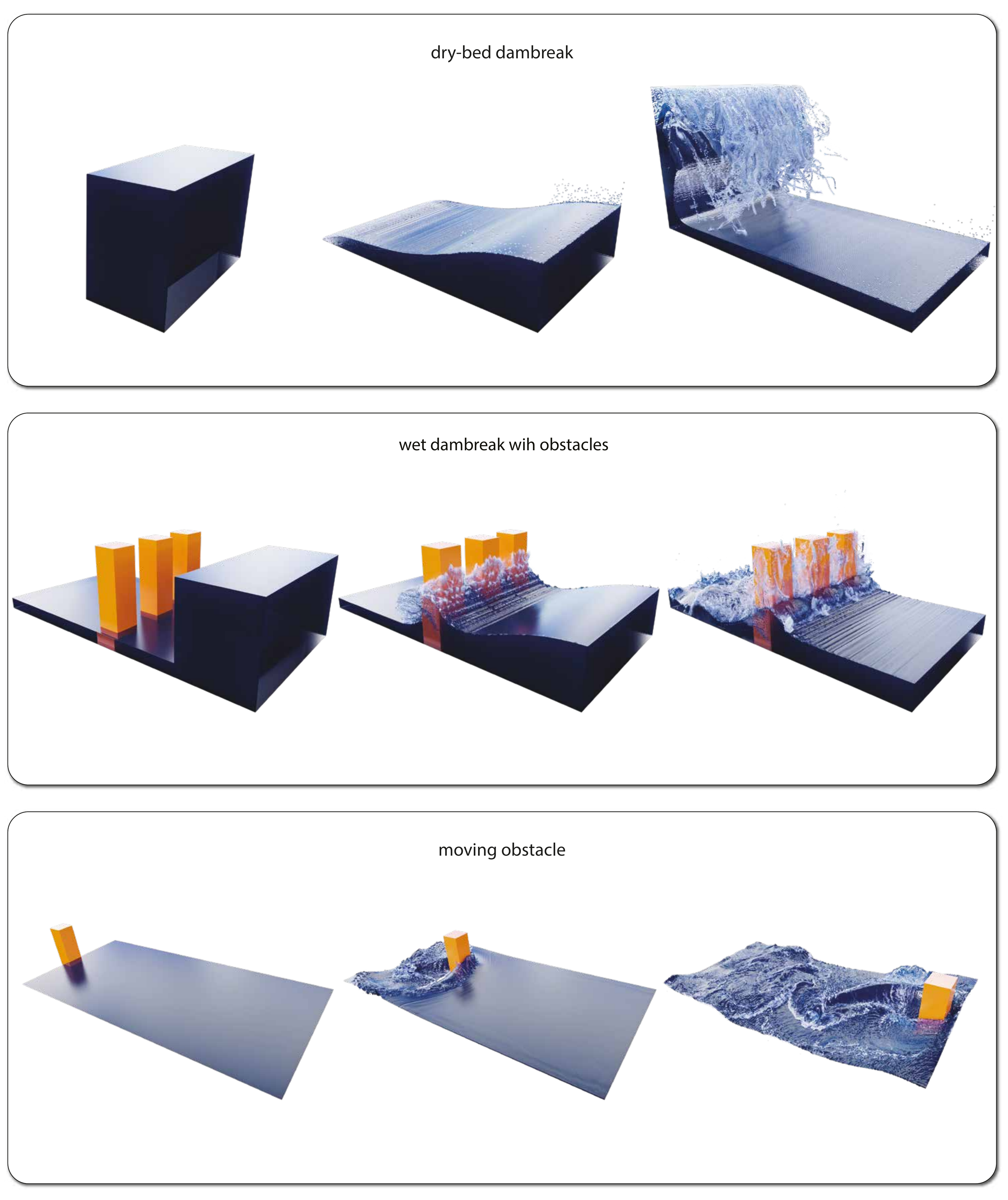}
\caption{\label{fig:weakness} Three example high-fidelity simulations where the multi-fidelity solver fails. The dry-bed dambreak consists of a column of fluid that is released at $t=0$, resulting in a wave impact on the domain boundary. The dambreak with obstacle uses the same initial fluid configuration as shown in figure \ref{fig:dambreakinit} but has solid obstacles (orange boxes) in the middle of the domain. The moving obstacle case consists of an obstacle (orange box) that is moving in a sinusoidal motion through a tank that is 40\% filled with fluid.}
\end{figure}

\noindent\textit{\textbf{Generalisation to other test cases}}\\
We showed that the neural network, which was trained on sloshing simulation data, is able to generalise to a wet dambreak problem. This is a promising result, but using the multi-fidelity solver for still more differing cases leads to unsatisfactory results. In figure \ref{fig:weakness} we show three example high-fidelity simulations where the multi-fidelity solver became unstable after only a few time steps.
These cases all have features that the neural network has never encountered before, i.e., a dry bed, static obstacles in the flow, and obstacles moving through the fluid. This is consistent with the results in \cite{ladicky_data-driven_2015,ladicky_physicsforests:_2017} where it is noted that generalisation is often an issue when using neural networks for fluid flow predictions. A carefully constructed training set may be needed to be able to generalise to a wide range of test cases.

\noindent\textit{\textbf{Physics interpretation of the multi-fidelity approach}}\\
The idea of our approach is to learn low-level features of the low-fidelity simulations and map these to the corresponding fine-grid counterpart. An in-depth analysis of why this approach works when using neural networks is challenging, due to the inherent non-linearities of the neural networks and the underlying physics.

\section{Conclusion}
We investigated the use of deep deconvolutional neural networks to enhance a low-fidelity PIC/FLIP fluid solver. An important novelty of our approach is that it is used intrusively in the low-fidelity solver and is able to enhance the solution at every time step. This is different from the common non-intrusive approaches that use neural networks either as a stand-alone solver or as a post-processing tool. The neural network in this paper was trained on randomly generated fluid sloshing data. After training, the neural network provided a significant increase in accuracy when compared to the low-fidelity solution, for the case of a sloshing simulation that was not comprised in the training data. Although the neural network introduced small oscillations on the fluid surface when considering a dambreak problem, it was still able to significantly enhance accuracy for this problem, which hints at the generality of our approach.

Our approach still has some issues. Firstly, the choice for the time step is rather restrictive, the time step needs to be kept small to keep the multi-fidelity solution stable. Secondly, our approach fails to enhance solutions that introduce new phenomena that were not encountered in the training set, e.g., obstacles in the flow. We expect that a carefully constructed training set may alleviate this issue and developing a procedure for this is scheduled for future work.

\bibliographystyle{ieeetr}
\bibliography{MyLibrary}

\begin{thebibliography}{10}

\bibitem{barreiro_smoothed_2013}
A.~Barreiro, A.~J.~C. Crespo, J.~M. Domínguez, and M.~Gómez-Gesteira,
  ``Smoothed particle hydrodynamics for coastal engineering problems,'' {\em
  Computers \& Structures}, vol.~120, pp.~96--106, Apr. 2013.

\bibitem{guo_convolutional_2016}
X.~Guo, W.~Li, and F.~Iorio, ``Convolutional neural networks for steady flow
  approximation,'' in {\em {ACM} {SIGKDD} {International} {Conference} on
  {Knowledge} {Discovery} and {Data} {Mining}}, {KDD} '16, (New York, USA),
  pp.~481--490, ACM, 2016.

\bibitem{mcadams_detail_2009}
A.~McAdams, A.~Selle, K.~Ward, E.~Sifakis, and J.~Teran, ``Detail preserving
  continuum simulation of straight hair,'' in {\em {ACM} {SIGGRAPH} 2009},
  {SIGGRAPH} '09, (New York, USA), pp.~62:1--62:6, ACM, 2009.

\bibitem{frost_moana:_2017}
B.~Frost, A.~Stomakhin, and H.~Narita, ``Moana: performing water,'' in {\em
  {ACM} {SIGGRAPH} 2017}, {SIGGRAPH} '17, (New York, USA), pp.~30:1--30:2, ACM,
  2017.

\bibitem{stomakhin_material_2013}
A.~Stomakhin, C.~Schroeder, L.~Chai, J.~Teran, and A.~Selle, ``A material point
  method for snow simulation,'' {\em ACM Trans. Graph.}, vol.~32,
  pp.~102:1--102:10, July 2013.

\bibitem{angelidis_controllable_2006}
A.~Angelidis, F.~Neyret, K.~Singh, and D.~Nowrouzezahrai, ``A controllable,
  fast and stable basis for vortex based smoke simulation,'' in {\em {ACM}
  {SIGGRAPH} 2006}, {SCA} '06, (Aire-la-Ville, Switzerland), pp.~25--32,
  Eurographics Association, 2006.

\bibitem{ravindran_reduced-order_2000}
S.~S. Ravindran, ``Reduced-order adaptive controllers for fluid flows {using}
  {POD},'' {\em Journal of Scientific Computing}, vol.~15, no.~4, pp.~457--478,
  2000.

\bibitem{audouze_reduced-order_2009}
C.~Audouze, F.~D. Vuyst, and P.~B. Nair, ``Reduced-order modeling of
  parameterized {PDEs} using time–space-parameter principal component
  analysis,'' {\em International Journal for Numerical Methods in Engineering},
  vol.~80, no.~8, pp.~1025--1057, 2009.

\bibitem{ladicky_data-driven_2015}
L.~Ladický, S.~Jeong, B.~Solenthaler, M.~Pollefeys, and M.~Gross,
  ``Data-driven fluid simulations using regression forests,'' {\em ACM Trans.
  Graph.}, vol.~34, no.~6, pp.~199:1--199:9, 2015.

\bibitem{ladicky_physicsforests:_2017}
L.~Ladický, S.~Jeong, N.~Bartolovic, M.~Pollefeys, and M.~Gross,
  ``Physicsforests: real-time fluid simulation using machine learning,'' in
  {\em {ACM} {SIGGRAPH} 2017}, pp.~22--22, 2017.

\bibitem{zhu_animating_2005}
Y.~Zhu and R.~Bridson, ``Animating sand as a fluid,'' in {\em {ACM} {SIGGRAPH}
  2005}, {SIGGRAPH} '05, (New York, USA), pp.~965--972, ACM, 2005.

\bibitem{bridson_fluid_2015}
R.~Bridson, {\em Fluid {Simulation} for {Computer} {Graphics}}.
\newblock CRC Press, 2015.

\bibitem{jiang_affine_2015}
C.~Jiang, C.~Schroeder, A.~Selle, J.~Teran, and A.~Stomakhin, ``The affine
  particle-in-cell method,'' {\em ACM Trans. Graph.}, vol.~34, no.~4,
  pp.~51:1--51:10, 2015.

\bibitem{phillips_interpolation_2003}
G.~M. Phillips, {\em Interpolation and {Approximation} by {Polynomials}}.
\newblock Springer-Verlag, 2003.

\bibitem{goodfellow_deep_2016}
I.~Goodfellow, Y.~Bengio, and A.~Courville, {\em Deep {Learning}}.
\newblock MIT Press, 2016.

\bibitem{aggarwal_neural_2018}
C.~C. Aggarwal, {\em Neural {Networks} and {Deep} {Learning}: {A} {Textbook}}.
\newblock Springer, 2018.

\bibitem{pinn1}
A.~D. Jagtap, E.~Kharazmi, and G.~E. Karniadakis, ``Conservative
  physics-informed neural networks on discrete domains for conservation laws:
  Applications to forward and inverse problems,'' {\em Computer Methods in
  Applied Mechanics and Engineering}, vol.~365, p.~113028, 2020.

\bibitem{pinn2}
Y.~Shin, J.~Darbon, and G.~E. Karniadakis, ``On the convergence and
  generalization of physics informed neural networks,'' 2020.

\bibitem{pinn3}
X.~Jin, S.~Cai, H.~Li, and G.~E. Karniadakis, ``N{S}fnets ({N}avier-{S}tokes
  flow nets): Physics-informed neural networks for the incompressible
  {N}avier-{S}tokes equations,'' 2020.

\bibitem{LES}
C.~Meneveau and P.~Sagaut, {\em Large Eddy Simulation for Incompressible Flows:
  An Introduction}.
\newblock Scientific Computation, Springer Berlin Heidelberg, 2006.

\bibitem{ibrahim_liquid_2005}
R.~A. Ibrahim, {\em Liquid {Sloshing} {Dynamics}: {Theory} and {Applications}}.
\newblock Cambridge University Press, 2005.

\bibitem{deconv}
V.~Dumoulin and F.~Visin, ``A guide to convolution arithmetic for deep
  learning,'' 2016.

\bibitem{deconv2}
A.~Radford, L.~Metz, and S.~Chintala, ``Unsupervised representation learning
  with deep convolutional generative adversarial networks,'' 2015.

\bibitem{clevert_fast_2015}
D.-A. Clevert, T.~Unterthiner, and S.~Hochreiter, ``Fast and accurate deep
  network learning by {Exponential} {Linear} {Units} ({ELUs}),'' {\em
  arXiv:1511.07289 [cs]}, Nov. 2015.
\newblock arXiv: 1511.07289.

\bibitem{maas_rectifier_2013}
A.~L. Maas, ``Rectifier nonlinearities improve neural network acoustic
  models,'' in {\em {ICML} {Workshop} on {Deep} {Learning} for {Audio},
  {Speech}, and {Language} {Processing}}, 2013.

\bibitem{abadi_tensorflow:_2016}
M.~A. et~al., ``{TensorFlow}: {A} system for large-scale machine learning,'' in
  {\em 12th {USENIX} {Symposium} on {Operating} {Systems} {Design} and
  {Implementation} ({OSDI} 16)}, pp.~265--283, 2016.

\bibitem{foam}
M.~Ihmsen, N.~Akinci, G.~Akinci, and M.~Teschner, ``Unified spray, foam and air
  bubbles for particle-based fluids,'' {\em The Visual Computer}, vol.~28,
  pp.~669--677, 06 2012.

\end{thebibliography}
\end{document}